 \documentstyle[11pt,aaspp4]{article}

 \slugcomment{To appear in Astrophys. J.}

\lefthead{KOHNO et al.}
\righthead{Dense Molecular Gas in the Star Forming Ring of NGC 6951}

\begin{document}

\title{Dense Molecular Gas Associated with the Circumnuclear Star Forming Ring
in the Barred Spiral Galaxy NGC 6951}

\author{Kotaro Kohno\altaffilmark{1}}
\affil{Department of Astronomy, School of Science, The University of Tokyo \\
       Bunkyo-ku, Tokyo, 113-0033, Japan}

\author{Ryohei Kawabe, and Baltasar Vila-Vilar\'o}
\affil{Nobeyama Radio Observatory \\
       Minamimaki, Minamisaku, Nagano, 384-1305, Japan}

\altaffiltext{1}{present address: Nobeyama Radio Observatory, 
                 e-mail: kotaro@nro.nao.ac.jp}

\begin{abstract}

We present high resolution (3$''$ -- 5$''$) observations of
CO(J=1$-$0) and HCN(J=1$-$0) emission from the circumnuclear
star forming ring in the barred spiral galaxy NGC 6951, a host of
a type-2 Seyfert nucleus, using the Nobeyama Millimeter Array 
and the Nobeyama 45 m telescope.

We find that the distribution of the HCN emission is different 
from that of CO in the circumnuclear region of NGC 6951; 
it is confirmed that CO emission is dominated 
by ``twin peaks'' morphology with two spiral arms, which are connected
to the dust lanes, as reported by Kenney et al. (1992).
On the other hand, although the HCN emission also shows 
a twin peaks morphology, the HCN peaks are spatially shifted downstream 
compared with the CO peaks.
Most of the HCN emission is associated with the circumnuclear ring, 
where vigorous star formation occurs. 
The HCN to CO integrated intensity ratio in the brightness temperature scale,
$R_{\rm HCN/CO}$, is also enhanced in the star forming ring. 
The peak value of the $R_{\rm HCN/CO}$ is about 0.16 -- 0.18, 
which is comparable to the $R_{\rm HCN/CO}$ in the starburst regions 
of NGC 253 and M82.
Consequently, the HCN emission spatially correlates better 
with the massive star forming regions
than the CO emission in the circumnuclear region of NGC 6951. 

The formation mechanism of dense molecular gas has been investigated.
No significant enhancement of $R_{\rm HCN/CO}$ is observed at the CO peaks,
which are interpreted as $x_1/x_2$ orbit crowding regions.
This suggests that the shocks along the orbit crowding do not promote
the formation of dense molecular gas effectively
but enhance the presence of low density GMCs in NGC 6951.
Instead, gravitational instability can account for the
dense molecular gas formation in the circumnuclear star forming ring
because Toomre's $Q$ value is below unity there.

The $R_{\rm HCN/CO}$ toward the type-2 Seyfert nucleus of NGC 6951
is 0.086 averaged over central $r$ $<$ 120 pc region.
This is a rather normal value compared with non-active galaxies 
such as the Milky Way,
and quite different from other type-2 Seyfert galaxies 
NGC 1068 and M51 where extremely high $R_{\rm HCN/CO}$ of $\sim 0.5$ 
have been reported.
The variety of $R_{\rm HCN/CO}$ values in these Seyfert nuclei 
would be attributed to the different physical conditions 
of the molecular gas around the nuclei.

\end{abstract}

\keywords{galaxies: individual (NGC 6951) --- galaxies: ISM --- galaxies: kinematics and dynamics --- galaxies: Seyfert --- galaxies: starburst}

\section{Introduction}

Observational studies of molecular gas in the Milky Way have clearly
shown that stars are formed from dense cores of molecular clouds
rather than their diffuse envelopes (e.g. \cite{lad92}). 
Hence it is essential to study the properties of dense molecular gas 
in order to understand star formation in galaxies, 
and particularly starburst phenomena.

Dense molecular material is investigated by observing the molecules
referred as ``high density tracers'', which require larger hydrogen
density than that of CO for their collisional excitation,
such as HCN, HCO$^+$, CS, and so on 
(e.g. \cite{mau89}; \cite{sss90}; \cite{hen91}; \cite{rie92}; 
\cite{sol92}; \cite{hab93}; \cite{aal95}; \cite{pag95}; \cite{pag97}). 
For example, CO(1$-$0) emission is excited even in low density
molecular gas ($n_{\rm H_2}$ $\sim$ 500 cm$^{\rm -3}$), 
whereas HCN(1$-$0) emission traces very dense molecular clouds 
($n_{\rm H_2}$ $>$ 10$^{\rm 4}$ cm$^{\rm -3}$)
due to its quite larger dipole moment
($\mu_{\rm HCN}$ = 3.0 Debye, whereas $\mu_{\rm CO}$ = 0.10 Debye).

Observational evidence suggesting a close relationship 
between dense molecular gas and massive star formation 
in various galaxies has been accumulated in recent years;
Solomon et al.\ (1992) reported a tight correlation between
the HCN(1$-$0) and far-infrared luminosity in 12 galaxies, suggesting that
the star formation rate is tightly related to the amount of 
the dense molecular gas measured by HCN emission.
This study has been expanded to larger samples (\cite{gao96}).
It has been also claimed that the HCN luminosity correlates 
with the H92$\alpha$ recombination line luminosity,
which is another tracer of massive star formation
(\cite{zha96}).
It is then natural to expect that dense molecular gas has 
a similar spatial distribution as that of the massive star forming regions.
Indeed, a rough association between dense molecular complexes
and star forming regions traced by radio continuum emission
has been observed in the prototypical starbursts 
M82 (\cite{sal95}; \cite{gol96}) and NGC 253 (\cite{ptj95}).
However, there are contradicting results on HCN -- FIR correlation
(e.g. \cite{aal95}), and only a few galaxies till present have been mapped 
in HCN emission with enough high spatial resolution 
to compare with the massive star forming regions.
It is therefore necessary to observe HCN emission from the galaxies 
which have extended massive star forming regions.

The formation mechanism of dense molecular clouds is another important issue.
At large spatial scales, gravitational instabilities of the molecular gas
(e.g. \cite{ken89}; \cite{elm94})
and the shocks associated with the orbit crowdings (e.g. \cite{cag85})
are thought to be responsible for the dense gas formation.
Closer to the centers of galaxies,
Inner Lindblad Resonances (ILRs) are considered to be essential
to explain the distribution and kinematics of molecular gas as well as
star formation (e.g. \cite{ken92}; \cite{tdw93}; \cite{sak95}),
and therefore dense gas cloud formation.

In order to address the issues described above, we observed 
CO(1$-$0) and HCN(1$-$0) emission in the central region of
a nearby barred spiral galaxy, NGC 6951,
using the Nobeyama Millimeter Array (NMA) and the Nobeyama 45 m telescope
\footnote{Nobeyama Radio Observatory is a branch
of the National Astronomical Observatory, an inter-university research
institute, operated by the Ministry of Education, Science, Sports
and Culture, Japan.}.
Optical and radio observations indicate a vigorous star formation 
in the circumnuclear region of this galaxy.
Both H$\alpha$ (\cite{mam93}; \cite{woz95}; \cite{roz96}; \cite{gon97}) 
and 6 cm radio continuum emission (\cite{vil90}; \cite{sai94}) 
are mainly concentrated in the central (a few hundred parsec) 
region of NGC 6951, indicating that most of the star formation occurs 
in a circumnuclear ring structure (Fig. \ref{fig1}).
The diameter of this circumnuclear star forming ring is 
about 6$''$ $\times$ 9$''$ (e.g. \cite{bar95}; \cite{bac93}). 
The star formation rate of the ring estimated from the H$\alpha$ luminosity
is $\sim$ 4 $M_\odot$ yr$^{\rm -1}$ after correction for internal extinction
(Table \ref{tbl1}), and is comparable to those in the central regions 
of nearby starburst galaxies, though the gas consumption timescale 
due to star formation is fairly longer than those in starbursts
(\cite{ken97}).
We can therefore resolve spatially the distribution of
dense molecular gas using a millimeter-wave interferometer,
and compare it with the extent of the vigorous star forming regions.
Near-infrared photometry has revealed the presence of
a large stellar bar with a semimajor radius of $\sim$ 44$''$
(\cite{mrk97}; \cite{fri96}; \cite{elm96}), which corresponds to $\sim$ 5 kpc
at the adopted distance of $D$ = 24.1 Mpc.
Optical images show two straight dust lanes along the bar 
(\cite{ken92}; \cite{mam93}; \cite{woz95}) 
which trace the shock front at the leading edges of the bar 
as predicted from numerical simulations (\cite{rha79}; \cite{ath92}).
CO(1$-$0) emission has been mapped with the Caltech millimeter array
(\cite{ken92}) and show a ``twin peaks'' morphology.
The positions of the ILRs deduced from their CO rotation curve indicate that
the two-armed CO peaks would be explained as a result of orbit crowding 
near the outer ILR (OILR).
Therefore, this circumnuclear ring of NGC 6951 provides us 
with an unique opportunity to investigate the relationship 
between the spatial variations of the physical conditions
of molecular gas and its dynamics, especially concerned orbit 
resonances and associated shocks. 

In addition to the study of the star forming activity,
in this object we can also address the relationship between an AGN
and molecular gas properties, because this galaxy hosts an active nucleus 
(type-2 Seyfert; \cite{bas93}; \cite{hfs95}). 
As it has been reported in the recent literature,
there seems to be a striking enhancement of 
the integrated HCN(1$-$0) to CO(1$-$0) intensity ratio
in brightness temperature scale (hereafter $R_{\rm HCN/CO}$)
toward the type-2 Seyfert nucleus of NGC 1068 
(\cite{jac93}; \cite{tac94}; \cite{hab95})
and the low-luminosity active galactic nucleus in M51 
(\cite{koh96}; \cite{mat98});
the $R_{\rm HCN/CO}$ is as high as $\sim$ 0.5 toward these two Seyfert nuclei.
These high $R_{\rm HCN/CO}$ values in a few hundred parsec scale region
have never been found anywhere in the Milky Way, implying extreme conditions 
of the molecular gas in the circumnuclear region of Seyfert galaxies. 
It is therefore quite exciting to explore 
whether such an extremely high $R_{\rm HCN/CO}$ is common or not
in the circumnuclear regions of Seyfert galaxies.

An H$\alpha$ + [NII] contour map (\cite{woz95}) superposed on 
an optical broad band image of NGC 6951 is shown in Fig. \ref{fig1},
together with the area observed with the NMA.
The properties of NGC 6951 were summarized in Table \ref{tbl1}.
In this paper we adopt a distance of NGC 6951 of 24.1 Mpc (Tully 1988),
which agrees well with that estimated using the Tully-Fisher relationship, 
23.4 Mpc (\cite{bot84}; see also \cite{mam93}).

\placetable{tbl1}
\placefigure{fig1}

We describe the observational parameters and results in Section 2,
and present new CO and HCN data of NGC 6951 in Section 3, which reveal
the different distribution of HCN emission from CO, namely,
the spatial variation of physical condition 
in the circumnuclear molecular clouds.
With these data, we discuss about the possible cause of the variation 
and its relationship to the central activity of NGC 6951 in Section 4. 
Finally we summarize the conclusions of this study in Section 5.

\section{Observations}

\subsection{Nobeyama Millimeter Array}

The central region of NGC 6951 was observed
in the J=1$-$0 line of CO and HCN with the NMA.
The NMA consists of six 10 m dishes providing 15 baselines simultaneously.
The observations were made during November 1995 to Feburary 1996
in the three available array configurations (AB, C, \& D).
Due to the limitation of the minimum projected baseline length (10 m),
extended structures larger than about 50$''$ in each channel map 
are not sampled in the observations.
The front-end are tunerless SIS receivers, 
whose receiver temperature is $\sim$ 50 K in double side band,
and the typical system noise temperature (in single side band)
were 300 -- 600 K for the HCN observations 
and 400 -- 800 K for CO the observations. 
A digital spectro-correlator FX was configured 
to cover 320 MHz with 1024 channels per baseline.
Sideband separation was achieved by 90$^\circ$ phase switching.
We observed a radio source 3C418 for 5 minutes
every 30 minutes in order to calibrate temporal variations of 
the visibility amplitude and phase. The passband across 1024 channels
was calibrated through observations of strong continuum sources 
3C273 and NRAO 530.
The flux density of the reference calibrator 3C418 was measured
at 88 GHz from comparisons with planets of known brightness temperatures.
The flux density of 3C418 at 114 GHz (CO observations) was
estimated by an extrapolation of its spectrum assuming a spectral index 
of -0.34 which was derived from the flux at 88 GHz 
and the lower frequency fluxes listed in the VLA calibrator database.
The uncertainty in the absolute flux scale is $\sim \pm$ 20 \%.

The raw data were calibrated and edited using the package UVPROC-II 
developed at NRO and Fourier transformed with natural {\it uv} weighting 
with the NRAO AIPS.
A conventional CLEAN method was applied to deconvolve the synthesized 
beam pattern.  We made CO channel maps with a synthesized beam
of 3$\farcs$9 $\times$ 3$\farcs$1 at a velocity width of 9.79 km s$^{\rm -1}$,
and HCN channel maps with a synthesized beam of
4$\farcs$7 $\times$ 4$\farcs$5 at 38.21 km s$^{\rm -1}$ velocity width 
every 19.11 km s$^{\rm -1}$ were made.
The typical noise levels of channel maps are 34 mJy beam$^{\rm -1}$ 
(= 260 mK in $T_{\rm b}$) for CO 
and 5.1 mJy beam$^{\rm -1}$ (= 37 mK) for the HCN data, respectively.
Parameters of the NMA observations described above are summarized 
in Table \ref{tbl2}.

\placetable{tbl2}

We searched continuum emission by binning-up line free channels of 
the HCN observations, but did not find any. 
No subtraction of continuum emission was therefore needed for the channel maps.
The 2 $\sigma$ upper limit of 
the 3.4 mm continuum emission is 2.4 mJy beam$^{\rm -1}$
(beam size 5$\farcs$4 $\times$ 5$\farcs$0 for this continuum search)
within the observed field. 

\subsection{NRO 45 m telescope}

In order to evaluate the missing flux of the interferometry,
we obtained single dish CO and HCN spectra toward the center of NGC 6951 
using the NRO 45 m telescope. The observations were carried out
in May 1996. In order to avoid the degradation of
the beam pattern, aperture efficiency, and pointing accuracy of the telescope, 
we used only the data taken when wind velocity was less than 5 m s$^{\rm -1}$.
The full half-power beam width, aperture efficiency,
and main-beam efficiency were measured to be
15$''$ (19$''$), 0.37 $\pm$ 0.02 (0.44 $\pm$ 0.02), 
and 0.45 $\pm$ 0.03 (0.50 $\pm$ 0.03) 
at the observed frequency of CO (HCN), respectively.
We used two cooled SIS mixer receivers equipped with sideband rejection
filters and observed both CO and HCN line simultaneously.
The beam squint of the two receivers had been aligned to less than 2$''$.
Absolute pointing of the antenna was checked every hour
using the SiO maser source T-Cep at 43 GHz, and was measured
as accurate as $\sim$ 2$''$ rms.
System noise temperature for CO and HCN observations were
typically 750 K and 350 K in single side band, respectively.
Sky emission were subtracted by position switching 
with two off-source positions
at offsets in azimuth of $\pm$ 5$\arcmin$ from the observed position.
Spectra of CO and HCN emission were obtained with 2048 channels
acousto-optical spectrometers of 250 MHz bandwidth, corresponding to 
a velocity coverage of 650 km s$^{\rm -1}$ for CO observations, 
and 850 km s$^{\rm -1}$ for HCN observations.
Only linear baselines were removed from the spectra and adjacent channels 
were binned to a velocity resolution of 10 km s$^{\rm -1}$ and 
20 km s$^{\rm -1}$ for CO and HCN spectra, respectively.
The resultant noise level of the CO spectrum was 9.5 mK rms
in $T^*_{\rm A}$ scale and 2.4 mK rms for the HCN spectrum.

\section{Results}

\subsection{Distribution of Molecular Gas}

\subsubsection{Channel maps, profile maps, and integrated intensity}

Fig. \ref{fig2} shows the channel maps of the CO emission and
HCN emission in the central 30$''$ $\times$ 30$''$ region 
(3.5 kpc $\times$ 3.5 kpc) of NGC 6951.
We detected significant ($>$ 3 $\sigma$) CO emission in 40 adjacent channels
with a velocity range of $V_{\rm LSR}$ = 1214 - 1594 km s$^{\rm -1}$. 
This velocity width (full width of zero intensity) of 390 km s$^{\rm -1}$ is 
almost the same as that of single-dish line profiles 
obtained with the FCRAO 14 m (\cite{you95}).
HCN emission was detected ($>$ 3 $\sigma$) over a velocity range of 
$V_{\rm LSR}$ = 1225 - 1567 km s$^{\rm -1}$, 
which is almost the same as that of the CO line.
Note that this is the first measurement of HCN emission from NGC 6951.
It is noteworthy that both CO and HCN emission are weak 
near the systemic velocity ($V_{\rm LSR} \sim 1428$ km s$^{-1}$; 
Table \ref{tbl4}), indicating that neither CO nor HCN emission show 
significant concentration toward the Seyfert nucleus of NGC 6951.

The velocity-integrated intensity maps of the CO and HCN emission are shown in
Fig. \ref{fig3}, together with the CO and HCN line profile maps.
These images were made by the calculation of the 0-th moment from 
the 3-dimensional data cube.
To minimize the contribution from noise, we computed these moment maps
with a clip level of 2 $\sigma$ for CO and 1.5 $\sigma$ for HCN 
in each channel maps. 

\placefigure{fig2}
\placefigure{fig3}

\subsubsection{45m observations and flux measurements}

Here we compare the CO and HCN flux from our NMA observations 
with single-dish observations
in order to evaluate the missing flux of our interferometry.
CO and HCN fluxes were measured by summing the emission from 
individual channel maps corrected for the primary beam attenuation.

The total CO flux within our field of view (65$''$) is 
334 $\pm$ 12 Jy km s$^{-1}$.
On the other hand, CO flux from the central 45$''$ diameter of NGC 6951 was 
measured to be 350 $\pm$ 41 Jy km s$^{-1}$ with the FCRAO 14 m telescope 
(\cite{you95}).
Because most of the CO emission in our NMA observations are confined within
the central 45$''$ diameter (i.e. beam size of the FCRAO 14 m), 
it seems that most of the single-dish flux is sampled in our NMA observations.
Note that low level CO emission has been detected over the disk of NGC 6951
(\cite{kun97}), and our interferometric observations may have missed
these weak extended emission.

We detected a total HCN flux of 18.7 $\pm$ 1.9 Jy km s$^{\rm -1}$ 
inside our field of view (84$''$).
Since there are no data for HCN flux observed with any 10 m class single-dish
telescope, we convolved our HCN data cube to 19$''$ resolution
and compared them with the 45 m spectrum.
Fig. \ref{fig4} shows the profiles of CO and HCN emission 
in main-beam temperature scale $T_{\rm MB}$ = $T^*_{\rm A}/\eta_{\rm MB}$.
We find that both the CO and HCN emission profiles from our 45 m observations 
agree very well with the convolved NMA profiles,
indicating that both CO and HCN maps from the NMA seems to recover 
almost all of single-dish flux again, and that we can safely discuss 
the distribution, kinematics, and line ratio of molecular gas below.
Note that the moment maps in the following figures recovers about 70 -- 90 \%
of single dish flux because these maps are made with a clipping 
of low-level emission.
All flux measurements, including line ratio calculation, are from
the maps without clippings.
Attenuation due to the primary beam pattern of 10 m dish
is also corrected for these flux measurements. 
The summary of 45 m observations are listed in Table \ref{tbl3}.

\placefigure{fig4}
\placetable{tbl3}

\subsubsection{Distribution of CO and HCN emission}

Fig. \ref{fig5} shows the spatial relationship between molecular gas, 
dust lanes, and the circumnuclear star forming ring traced by H$\alpha$ 
emission (\cite{woz95}) and
the 6 cm radio continuum emission (\cite{sai94}).
CO emission in the central region of NGC 6951 is confirmed 
to be dominated by two strong peaks on spiral arms (hereafter the CO arms),
referred as ``twin-peaks'' by Kenney et al. (1992), which are directly
connected to the dust lanes along the bar. 
The inner part of the two CO arms corresponds to
the circumnuclear star forming ring.
These features agree well with the previously reported one (\cite{ken92}).
On the other hand, as it can be clearly seen in Fig. \ref{fig5},
{\it the distribution of the HCN emission is significantly different 
to that of CO}.
First, though the HCN emission also shows a twin peak morphology, 
the peaks are spatially shifted from those of CO.
If we assume that the spiral arms of NGC 6951 seen in optical images 
(Fig. \ref{fig1}) are trailing as in case of almost all spiral galaxies,
the rotation is clockwise as indicated by an arrow in the figure 
and the HCN emission is shifted downstream. 
Moreover, the HCN peaks are located closer to the nucleus than the CO peaks.
About 44 \% of the total CO flux comes from the ring ($r < 6''$), 
whereas 60 \% of the HCN emission is associated with it.
Namely, {\it the HCN emission correlates better spatially
with the massive star forming region than the CO emission in NGC 6951}.
Although the presence of a strong and non-uniform extinction is reported from 
the H$\alpha$/H$\beta$ ratio variations across the ring (\cite{bar95}),
both H$\alpha$ and radio continuum show good agreement with the HCN emission.

\placefigure{fig5}

Association between the HCN distribution and the massive star forming regions 
has been reported in prototypical starburst galaxies M 82 (\cite{sal95}; 
Golla et al.\ 1996) and NGC 253 (Paglione et al.\ 1995), although these are 
edge-on galaxies.  
Dense molecular gas is also distributed in the starburst ring (or arms) of 
NGC 1068 (\cite{jac93}; \cite{hab95}). 
Particularly, sensitive observations with the PdB interferometer have 
clearly depicted the HCN emission associated with the starburst ring of 
NGC 1068 (\cite{tac94}), which is quite similar to the case of the 
circumnuclear star forming ring of NGC 6951.  
The HCN distribution of IC 342 (\cite{dow92}) 
may also be an example of such coincidence
\footnote{Although Downes et al.\ (1992) states that neither CO nor HCN trace 
star formation in IC 342, the HCN emission mostly arises from a central ring 
where massive star formation occurs (\cite{wri93}; Telesco et al.\ 1993; 
\cite{bfg97}). On the other hand, a considerable fraction of the CO emission 
comes from the two bright ridges which extend for $\sim$ 30$''$ from the 
nucleus (\cite{ish90}; see also the CO mosaic map in \cite{wri93}).
Accordingly, the HCN emission may correlate better spatially with the star 
formation than the CO emission in IC 342, as in the case of NGC 6951.}.

We note that the FIR/HCN luminosity ratio $L_{\rm FIR}/L'_{\rm HCN}$ in 
NGC 6951 is about 400 (Table \ref{tbl2}), which is similar to the values 
in ultra-luminous IR galaxies,
whereas $L_{\rm FIR}/L'_{\rm CO}$ $\sim$ 40 is about 5 times smaller 
than those in ultra-luminous IR galaxies (Solomon et al.\ 1992).
These comparisons support the results of Solomon et al. (1992),
which have shown that $L_{\rm FIR}/L'_{\rm HCN}$ is
independent of the $L_{\rm FIR}$, whereas $L_{\rm FIR}/L'_{\rm CO}$ 
increases with $L_{\rm FIR}$.

Interestingly, there is no central peak in HCN emission. 
This is different from the other Seyfert galaxies NGC 1068 and M51, 
where there is a strong concentration of HCN emission toward the nucleus, 
in spite of weak or no CO peak there 
(\cite{jac93}; \cite{tac94}; \cite{hab95}; \cite{koh96}).

\subsubsection{Molecular gas mass and surface density}

The mass of molecular gas is estimated from the CO emission
assuming the Galactic $I$(CO) to $N$(H$_{2}$) conversion factor.
Adopting a conversion factor of 
$X_{\rm CO} = 3 \times 10^{\rm 20}$ cm$^{\rm -2}$ 
(K km s$^{\rm -1}$)$^{\rm -1}$ (\cite{sco87}; \cite{sol87}),
the mass of molecular hydrogen is 
\begin{equation}
M_{\rm H_2}
= 1.2 \times 10^{4} \times 
  \left(\frac{S({\rm CO)}}{\mbox{Jy km s$^{\rm -1}$}}\right) 
  \left(\frac{D}{\mbox{Mpc}}\right)^{2} M_\odot\mbox{,}
\end{equation}
where $S({\rm CO})$ is the CO flux density.
Taking into account that the mass of molecular contents including He 
and other elements as $M_{\rm gas} = 1.36 \times M_{\rm H_2}$,
the total mass of molecular gas detected in our integrated intensity map
is $3.2 \times 10^{\rm 9}$ $M_\odot$, and the molecular gas mass
within $r < 6''$, the radius of the circumnuclear ring,
is $1.4 \times 10^{\rm 9}$ $M_\odot$.

The surface mass density of molecular hydrogen on the galaxy plane is 
calculated as
\begin{equation}
\Sigma_{\rm H_2} 
= 4.8 \times \cos(i) \times 
  \left( \frac{I({\rm CO})}{\mbox{K km s$^{-1}$}} \right)
  M_\odot \mbox{\ pc$^{-2}$},
\end{equation}
where $i$ is the inclination of the galaxy ($i$ = 42$^\circ$ for NGC 6951). 
In case of our CO observations, 
the 1$\sigma$ level of CO integrated intensity map,
1.9 Jy beam$^{\rm -1}$ km s$^{\rm -1}$ or 15 K km s$^{-1}$, corresponds to
a face-on molecular gas surface density 
$\Sigma_{\rm gas} = 1.36 \times \Sigma_{\rm H_2}$ 
= 73 M$_\odot$ pc$^{\rm -2}$, and the CO peak (27 $\sigma$)
is $2.0 \times 10^3$ M$_\odot$ pc$^{\rm -2}$.
Note that this gas surface density does not depend on the adopted distance 
to the galaxy.

\subsection{$R_{\rm HCN/CO}$ ratio}

CO(1$-$0) emission is excited even in low density
molecular gas ($n_{\rm H_2}$ $\sim$ 500 cm$^{\rm -3}$)
whereas HCN(1$-$0) emission traces very dense molecular clouds 
($n_{\rm H_2}$ $>$ 10$^{\rm 4}$ cm$^{\rm -3}$).
Therefore the comparison of CO and HCN intensity is a measure of gas density 
if both CO and HCN emission come from the same volume. 
Fig. \ref{fig6} shows the dependence of $R_{\rm HCN/CO}$ on kinetic temperature 
and density of molecular gas (\cite{koh98}), 
which was computed employing the large velocity gradient approximation
(\cite{sas74}; \cite{gak74})
\footnote{$R_{\rm HCN/CO}$ values are often interpreted as a measure 
of thermal gas pressure (\cite{hab97b}; \cite{pag98}). 
However, gas pressure is expressed as a product of gas density and 
kinematic temperature, and is strongly dependent on the gas density 
(Fig. \ref{fig6}; see also \cite{pag98})}.
In case of external galaxies, observing beams are usually too large 
to distinguish the individual molecular cloud structures, and the one-zone 
assumption is no longer valid. 
Thus CO and HCN line intensity ratio is interpreted as the fraction of 
dense molecular gas to the total (including diffuse) molecular gas 
within the observing beam.

\placefigure{fig6}

\subsubsection{Distribution of $R_{\rm HCN/CO}$}

We convolved the CO map to the same beam size as that of the HCN map
in order to compare directly the HCN and CO line intensities.
Fig. \ref{fig7}(a) is a map of the HCN to CO integrated intensity ratio
in brightness temperature scale ($R_{\rm HCN/CO}$),
and Fig. \ref{fig7}(b) is a comparison of the $R_{\rm HCN/CO}$ 
with the 6 cm continuum emission (\cite{sai94}).
It is immediately evident that {\it $R_{\rm HCN/CO}$ values are significantly 
enhanced in the 6 cm continuum peaks, where intense star formation occur}.
The highest value of $R_{\rm HCN/CO}$ is about 0.16 -- 0.18, which is 
similar to the ratios observed in the molecular gas in the nuclear starburst 
region of M82 and NGC 253 (\cite{sal95}; Paglione et al.\ 1995).
We have also detected strong HCN emission toward the nucleus of
the young starburst galaxy NGC 3504, where $R_{\rm HCN/CO}$ exceeds 0.2
(\cite{koh97}).
The Sgr-B2 cloud, a massive star forming region in the central region 
of the Milky Way, also has a very high $R_{\rm HCN/CO}$ of $\sim$ 0.2 
(\cite{jac96}). 
On the other hand, an $R_{\rm HCN/CO}$ of 0.16 -- 0.18 is much larger than
that observed in the Galactic disk, $R_{\rm HCN/CO}$ $\sim$ 0.026 $\pm$ 0.008 
(\cite{hab97b}) and the central regions of ``normal'' galactic nuclei
(which means the nuclei showing no extreme activities),
such as Milky Way where a ratio $\sim$ 0.08 is observed
averaged over the central 600 pc region (\cite{jac96}; \cite{hab97b}).

\placefigure{fig7}

Fig. \ref{fig8} shows the azimuthally averaged radial profile of 
$R_{\rm HCN/CO}$.  Again, we can see that the ratio is enhanced 
at the circumnuclear star forming ring.
The azimuthly averaged value of the $R_{\rm HCN/CO}$ 
in the circumnuclear ring is 0.13.
This is comparable to the azimuthally averaged $R_{\rm HCN/CO}$ 
in the Starburst ring of NGC 1068,
which is also as high as 0.1 (\cite{hab95}).

\placefigure{fig8}

Turning now to the nucleus, we see that the $R_{\rm HCN/CO}$ 
toward the type 2 Seyfert nucleus of NGC 6951 is 0.083 averaged over 
the central $r < 120$ pc region (Fig. \ref{fig8}).
This ratio is a rather normal value compared with other non-active galaxies,
and quite different from the previously studied Seyfert galaxies
NGC 1068 and M51, where extremely high $R_{\rm HCN/CO}$ of $\sim$ 0.5 have been
reported (\cite{jac93}; \cite{tac94}; \cite{koh96}; \cite{mat98}).
We note that the lack of a central peak in $R_{\rm HCN/CO}$ radial distribution
is in contrast with the other galaxies;
high resolution measurements of $R_{\rm HCN/CO}$ 
in NGC 1068 (\cite{hab95}), M51 (\cite{koh96})
and NGC 6946 (Helfer \& Blitz 1997a) have been shown 
centrally peaked $R_{\rm HCN/CO}$ radial profile.

As a summary of this section,
we present the CO, HCN, and $R_{\rm HCN/CO}$ distribution
in a polar coordinate graph. Fig. \ref{fig9} shows the definition of the axis 
used to measure the radius $R$ and the position angle $\phi$, 
and Fig. \ref{fig10} shows the maps.
The features described above are also clearly revealed in these maps; 
the HCN emission and $R_{\rm HCN/CO}$ peaks are shifted downstream 
with respect to the CO peaks.
The difference of the position angle between the CO peaks 
and the $R_{\rm HCN/CO}$ peaks is $\sim$ 50$^\circ$ -- 60$^\circ$.
Most of the HCN emission is associated with the star forming ring 
at a radius of $3'' \sim 6''$,
whereas a considerable fraction of the CO emission 
originates from the two arms located beyond the ring ($r > 6''$).
The peaks of $R_{\rm HCN/CO}$ are coincided with 
the massive star formation traced by the 6 cm radio continuum emission,
suggesting a tight relationship between dense molecular gas and 
vigorous star formation in NGC 6951.

\placefigure{fig9}
\placefigure{fig10}

\subsubsection{Mass fraction of dense molecular gas}

It may be meaningful to interprete $R_{\rm HCN/CO}$ values 
as the mass fraction of dense molecular gas,
$M_{\rm dense}/M_{\rm total}$,
where $M_{\rm dense}$ is the mass of dense molecular gas traced by HCN(1$-$0),
and $M_{\rm total}$ is the total mass (i.e. both dense and diffuse)
of molecular gas traced by CO.
$M_{\rm dense}$ can be estimated using an ``HCN conversion factor'', 
$X_{\rm HCN} \equiv N({\rm H_2})/L'({\rm HCN})$.
$X_{\rm HCN}$ is computed to be 
$\sim 20$ $M_\odot$ (K km$^{-1}$ pc$^2$)$^{-1}$ or 
$1.3 \times 10^{21}$ cm$^{-2}$ (K km s$^{-1}$)$^{-1}$,
assuming that the HCN emission traces gas in gravitationally bound or 
virialized clouds (Solomon et al.\ 1992).
Together with the adopted CO conversion factor here,
$X_{\rm CO} = 3 \times 10^{20}$ cm$^{-2}$ (K km s$^{-1}$)$^{-1}$,
we find that $R_{\rm HCN/CO} = 0.1$ corresponds to 
$M_{\rm dense}/M_{\rm total} = 43$ \%.
Accordingly, in starburst regions where high $R_{\rm HCN/CO}$ values of 
$\sim$ 0.2 are observed,
$M_{\rm dense}/M_{\rm total}$ values may be close to $\sim$ 100 \%,
i.e. the molecular gas mass is mostly dominated by the dense component.
In the center of M82, a high fraction of dense molecular gas 
($M_{\rm dense}/M_{\rm total} \sim 40$ \%) has been estimated
based on HCN (\cite{bas93}) and high-J CO (\cite{gus93}) observations
\footnote{We would like to make a brief comment on the extremely high 
$R_{\rm HCN/CO}$ values of $\sim$ 0.5 toward some Seyfert nuclei such as 
NGC 1068 and M51. 
If we adopt the same $X_{\rm HCN}$ in these regions, 
it immediately results in an overestimation of dense molecular gas mass 
because $M_{\rm dense}/M_{\rm total}$ exceeds 100 \%.
We suggest that $X_{\rm HCN}$ may have a significantly different value there;
Kohno (1998) has examined the dependence of the $X_{\rm HCN}$ on the physical 
properties of gas based on the LVG model (\cite{sak97}) and shown that 
$X_{\rm HCN}$ tends to increase under very high kinematical temperature 
($T_{\rm kin} > 100$ K) and high gas density 
($n_{\rm H_2} \sim 10^5$ cm$^{-3}$) conditions.
In fact, $T_{\rm kin}$ and $n_{\rm H_2}$ have been estimated to be $\sim 100$ K 
and $\sim 10^5$ cm$^{-3}$, respectively, in the central regions of NGC 1068 
(\cite{tac94}) and M51 (\cite{mat98}).
Anyway, essentially {\it all} of the molecular gas must be dense
in regions where very high $R_{\rm HCN/CO}$ values are observed.
}.
Note that the $M_{\rm dense}/M_{\rm total}$ values derived here could
contain considerable errors mainly due to the uncertainties 
of both $X_{\rm HCN}$ and $X_{\rm CO}$.

\subsection{Kinematics of Molecular Gas}

\subsubsection{Velocity field}

Fig. \ref{fig11} shows an intensity-weighted isovelocity contour map
of the CO emission. This map was made by computing the first moment of 
the CO data cube, as $<\!\!v\!\!>\,\, = \Sigma v_i S_i/\Sigma S_i$.
Circular motion dominates the kinematics in the central region of NGC 6951,
however, we can see a ``S-shape'' twist of isovelocity contours
near the systemic velocity, suggesting the existence of non-circular motions.

\placefigure{fig11}

We determined the kinematical parameters of the circumnuclear molecular 
gas ring (dynamical center, position angle of the major axis, 
inclination angle, and systemic velocity) by a least square fitting 
of the intensity-weighted isovelocity field as a circular rotation.
We used a fitting only to the central region ($1'' < r < 5''$) because
the gas distribution is far from axisymmetric in the outer parts ($r < 6''$)
and also large deviations from circular motion appear to exist there.
The GAL package in AIPS was used for this analysis.
The derived parameters are listed in Table \ref{tbl4}.
Note that these quantities may still contain systematic errors 
due to the presence of non-circular motions seen in the mean velocity map.

\placetable{tbl4}

The position of the dynamical center determined from this observations roughly 
corresponds with the position of the Seyfert nucleus defined 
as the central peak of the 6 cm radio continuum
(\cite{sai94}) within a difference smaller than 0$\farcs$5.
The systemic velocity is consistent with the H$\alpha$ observations,
$V_{\rm hel}$ = 1418 $\pm$ 15 km s$^{\rm -1}$ (\cite{mam93}).
The position angles of the major axis estimated from optical isophotal fits 
range from 138$^\circ$ (\cite{mam93}) to 170$^\circ$ (RC3).
The derived value from our CO data, $P.A.$ of 130$^\circ$, favors the smallest one.
The inclination angle was highly uncertain from the fitting, 
and we adopted the value determined by M\'arquez \& Moles (1993).

Position - velocity diagrams ($pv$ diagrams)
along the major axis ($P.A.$ = 130$^\circ$)
of NGC 6951 are shown in Fig. \ref{fig12}.
It is confirmed that the CO emission is weak toward the nucleus in
the $pv$ diagram along the major axis.

\placefigure{fig12}

\subsubsection{Rotation curve}

Fig. \ref{fig12}(a) shows the rotation curve of the inner part of NGC 6951
obtained by averaging the intensity-weighted mean velocities (Fig. \ref{fig11})
within $\pm 5^\circ$ area of the major axis.
It should be noted that the velocity field contains non-circular motions
and the derived rotation curve must be treated as an averaged rotation curve,
possibly with significant local deviations.

\placefigure{fig12}

The dynamical mass within a radius $r$ is calculated as
\begin{equation}
M_{\rm dyn}
= \frac{r v^2(r)}{G} 
= 2.3 \times 10^5 \times \left( \frac{r}{\mbox{kpc}} \right) \left( \frac{v(r)}{\mbox{km s$^{\rm -1}$}} \right)^2 
M_\odot,
\end{equation}
assuming a spherical mass distribution and Keplerian rotation.
The dynamical mass within the radius of $6''$ is then calculated as 
$4.9 \times 10^{\rm 9}$ $M_\odot$,
so the ratio of dynamical mass to the total molecular gas mass 
$M_{\rm gas}/M_{\rm dyn}$ is about 29 \% within the ring.
This is a very high value, yet such large gas mass fraction 
is often observed in the central regions of galaxies (e.g. \cite{tur94}).

The angular velocity $\Omega(r) = v(r)/r$, epicyclic frequency 
$\kappa(r) = [2 \{ v(r)/r \} \{ v(r)/r + dv(r)/dr \} ]^{0.5}$, 
and $\Omega-\kappa/2$ calculated from the rotation curve, 
are shown in Fig. \ref{fig13}(b). 
ILRs will exist wherever $\Omega-\kappa/2$ exceeds the
pattern speed of the bar $\Omega_{\rm bar}$ at some location within corotation.
The bar pattern speed (indicated as dashed line) was estimated to be 
$\sim$ 50 km s$^{\rm -1}$ kpc$^{\rm -1}$,
assuming that corotation is located at the end of the bar. 
We took the bar radius as 44$''$ from Mulchaey et al. (1997),
and the rotation velocity at the bar end was derived from 
the H$\alpha$ data (\cite{mam93}).
It is evident that the bar pattern speed remains below 
$\Omega-\kappa/2$ within a radius of $\sim$ 800 pc.
This is similar to the loci of the CO arms, $r \sim 6''-10''$,
and roughly agrees with the analysis by Kenney et al. (1992).
The inner ILR (IILR) should be located near the nucleus, 
however, we cannot elaborate on the location of the IILR accurately 
because our rotation curve in Fig. \ref{fig13}(a) is affected 
by limited angular resolution in the central parts (within $r \sim 400$ pc).

\subsubsection{Deviation from circular rotation}

Since deviations from pure circular motion can be more clearly seen 
in velocity residual maps, a model velocity field
derived from the rotation curve in Fig. \ref{fig13}(a) assuming
pure circular motion was subtracted from the observed velocity field
(Fig. \ref{fig11}).
The resultant velocity residual map is shown
in Fig. \ref{fig14}, overlaid on the CO intensity map (greyscale)
and the position of the dust lanes (dashed lines).

\placefigure{fig14}

This comparison shows that the largest deviations
from circular motion occur at the end of the two CO arms 
which are directly connected to the dust lanes.
If we assume that the spiral arms in NGC 6951 are trailing,
then these dust lanes are located at the leading sides
of the bar, and the southwestern part of the galaxy is the near side,
indicating that the deviations associated with the
dust lanes seen in Fig. \ref{fig14} could be radial inflowing motion
along the bar, as predicted by theory
(Roberts et al.\ 1979; \cite{ath92}).
This feature has been reported in barred spiral galaxies such as
NGC 4314 (\cite{bsk96}) and NGC 1530 (\cite{rad97}).
Inside the ring, the spiral dust lane pattern has been reported
(\cite{bar95}) and it is also important to know whether the gas
near the Seyfert nucleus of NGC 6951 is inflowing or outflowing.
Some models, beyond the scope of this paper,
are necessary to make further interpretation of the velocity field
and to derive meaningful quantities such as net mass infall rate.

\subsubsection{Velocity dispersion}

Fig. \ref{fig15} shows the intensity-weighted CO velocity dispersion
along the line of sight, as derived from the second moment of the
CO data cube, $\sigma_v = (\Sigma (v_i - <\!\!v\!\!>)^2 S_i)/(\Sigma S_i)$.
This map shows the significant enhancement of the second moment value
near the two CO peaks.
The large velocity width near the CO peaks is also seen 
in the CO profile maps, Fig. \ref{fig3}(c). 
The southern CO peak is much weaker than the northern CO peak,
but also shows a broad (FWZI $\sim$ 200 km s$^{-1}$) line profile.
Kenney (1996) also has pointed out that CO spectra near the northern CO peak
show broad and complex line profiles in this galaxy.
It should be noted, however, that this map contains both the intrinsic velocity
dispersion of the gas and the gradient of rotation velocity in the observing beam.
For instance, the very large velocity dispersion area along the minor axis of
the galaxy seen in Fig. \ref{fig15} represents the steep rise of the rotation curve 
in the central region of the galaxy. 
In order to estimate the intrinsic gas velocity dispersion $\sigma_v$,
we subtracted the velocity component of the systematic velocity gradient within
observing beam, $\sigma_{\rm rot}$, from the observed velocity dispersion, 
$\sigma_{\rm obs}$,
as $\sigma_v = \sqrt{(\sigma_{\rm obs})^2 - (\sigma_{\rm rot})^2}$.
A circular motion described by the rotation curve in Fig. \ref{fig13} was assumed
as a model velocity field in the subtraction.
We find that gas the velocity dispersion is as high as $\sim$ 50 km s$^{\rm -1}$
in the Northern CO peak, and $\sim$ 30 km s$^{\rm -1}$ in the Southern CO peak.
On the other hand, the velocity dispersion at the ring is in the range
of 10 - 20 km s$^{\rm -1}$.
Note that these values are still upper limits to the true gas velocity dispersions
because we don't take into account the non-circular motions here.
Nevertheless, it is likely that the gas velocity dispersions at the CO peaks
is significantly higher than that in the circumnuclear ring because shocks associated
with the crowding of gas stream lines could keep the gas 
in a rather turbulent state there.
Therefore, we speculate that the velocity dispersion in the CO peaks 
may be significantly larger than that in the circumnuclear ring.

\placefigure{fig15}

\section{Discussion}

The CO and HCN data presented in the previous section revealed that
the distribution of HCN emission is different from that of CO,
in other words, there exists a spatial variation of the molecular line ratio
$R_{\rm HCN/CO}$.
In this section, we discuss about the relationship between 
the observed properties of molecular gas and activities,
namely, the circumnuclear star formation and Seyfert nucleus.

\subsection{Dense Molecular Gas Associated with the Circumnuclear Star Forming Ring}

As seen in the previous section, the distribution of dense molecular gas
correlates very well with the massive star forming region 
in the center of NGC 6951, and the $R_{\rm HCN/CO}$ ratio is also enhanced 
at the circumnuclear ring where vigorous star formation occurs.
A spatial coincidence of the HCN cloud complex and 
nuclear massive star forming regions, and an enhancement of 
$R_{\rm HCN/CO}$ there, have been found in the central regions 
of other star forming galaxies, as listed in the previous section.
These spatial coincidence between dense molecular gas and 
massive star forming regions is consistent 
with the quantitative correlation between HCN emission 
and tracers of massive star formation
(\cite{sol92}; \cite{gao96}; \cite{zha96}; \cite{pag97}).
Considering the fact that stars are formed in the {\it dense molecular cores}
rather than the diffuse envelopes of the molecular clouds 
in the Milky Way (e.g. \cite{lad92}),
these quantitative and spatial correlation seem to indicate that 
the presence of dense molecular gas in these a few hundred parsec scale 
in the central regions of galaxies could be a {\it cause} of massive star formation.
Therefore, it is now essential to address the formation mechanism 
of dense molecular clouds in order to explain the varieties 
of star formation in the central regions of galaxies. 
 
What kind of physical processes govern the spatial variation 
of $R_{\rm HCN/CO}$, namely, the distribution of dense molecular gas
in the circumnuclear region of NGC 6951?
Several mechanisms have been proposed to explain dense molecular gas cloud formation 
in the circumnuclear rings within the ILRs. 
One is the shocks along the orbit crowding; a crowding of gas stream lines will cause 
frequent collisions of the molecular gas clouds and shocks there, which can eventually
form dense molecular gas (\cite{cag85}). 
Another scenario is the gravitational instability of the accumulated gas 
near the ILRs (\cite{elm94}).
In order to address this issue, 
we first compared the distribution of $R_{\rm HCN/CO}$ with the maps of
the total gas surface density and the gas velocity dispersion,
as shown in Fig. \ref{fig16}.
In this comparison, we see the following two features, 
(i) in spite of the significant CO concentration,
$R_{\rm HCN/CO}$ is {\it not} enhanced in the crowding regions of $x1/x2$ orbits,
and (ii) there exists a possible anti-correlation between $R_{\rm HCN/CO}$ 
and the gas velocity dispersion.
Any scenarios to explain dense molecular gas formation should 
also account for these observed features.
In the following section, we will discuss about the several possible scenarios.

\placefigure{fig16}

\subsubsection{Shocks along the orbit crowding}

It appears to be evident that a dense cloud formation scenario by shocks 
along the orbit crowding can not explain the observed feature (i).
Dust lanes and the molecular gas spirals located at the leading edges
of stellar bars are considered to trace shocks associated with
the orbit crowding of the distinct gas orbit families, 
$x_1$ and $x_2$ (e.g. \cite{ath92}).
In the case of NGC 6951, we find a possible infalling motion 
along the bar, and an enhancement of the CO velocity dispersion there, 
which may support the interpretations since these agree well 
with the theoretical predictions.
We therefore conclude that the shocked regions along the leading edges of the bar
does not promote the formation of {\it dense} molecular gas efficiently
but enhance the presence of low density GMCs.

Similar situations can be found in the central regions of other galaxies;
in the center of IC 342, a ring-like structure with two offset ridges of molecular gas 
is observed in CO emission (\cite{ish90}), 
while HCN emission is detected only in the central ring-like region (\cite{dow92}).
The ring-like gas distribution and two offset gas ridges are due to ILRs
with a stellar bar (\cite{bfg97} and references therein).
High CO brightness temperature, large gas velocity dispersion, 
and deviations from circular motions observed in the
offset CO ridges strongly suggest the presence of shocks there (\cite{ish90}), 
however, the $R_{\rm HCN/CO}$ ratio is as small as $\sim$ 0.05 
compared with the central ring region where $R_{\rm HCN/CO}$ values are enhanced 
to $\sim$ 0.12 (\cite{dow92}). 
This is also the case in another barred spiral galaxy NGC 1530 (\cite{rad97}).
As pointed out by Downes et al. (1996) and Reynaud \& Downes (1997), 
these shocks along the orbit crowding regions
may be too turbulent to form dense molecular gas 
and the dense molecular gas fraction remains rather low there.
This would account for the low star formation efficiencies (SFEs) 
of the molecular gas along the dust lanes in the bar
because the dense molecular gas fraction is considered 
to correlate with the SFEs (Solomon et al.\ 1992).

\subsubsection{Gravitational instability}

Another possibility is gravitational instabilities of the molecular gas.
Especially, the observed feature (ii), that is a possible anti-correlation 
between the distribution of $R_{\rm HCN/CO}$ and
the gas velocity dispersion, 
implies that the large velocity dispersion in the orbit crowding region 
prevents the molecular gas from collapsing into the dense molecular clumps 
through the gravitational instabilities of the gas.

In order to check whether gravitational theory can successfully explain
the spatial variation of the $R_{\rm HCN/CO}$ or not, we estimated the Toomre
$Q$ parameter (Toomre 1964) in the circumnuclear region of NGC 6951.
The Toomre $Q$ parameter characterizes the criteria for local stability
in thin isothermal disks and is expressed as 
$Q = \Sigma_{\rm crit}/\Sigma_{\rm gas}$, where $\Sigma_{\rm gas}$ is 
the molecular gas surface density $\Sigma_{\rm H_2 + He}$, 
and $\Sigma_{\rm crit}$ is the critical surface density, given by
\begin{equation}
\Sigma_{\rm crit} = \frac{\sigma_r \kappa}{\pi G},
\end{equation}
where $\sigma_r$ is the velocity dispersion in the radial direction, and
$\kappa$ is the epicyclic frequency. 
If the gas surface density $\Sigma_{\rm gas}$
exceeds the critical surface density $\Sigma_{\rm crit}$, then
an uniform gas disk is unstable to form rings or clumps which can ultimately
collapse into dense molecular gas fragments and form stars.
Here we assume an isotropic velocity dispersion, namely, the velocity 
dispersion in the radial direction $\sigma_r$ is the same as 
the one-dimensional velocity dispersion $\sigma_v$.
At the radius of the ring, $r=4''$ or 470 pc, we find $\Sigma_{\rm crit}$ is
\begin{equation}
\Sigma_{\rm crit} = 
630 \times \left(\frac{\sigma_v}{15 \mbox{\ km s$^{-1}$}}\right) 
           \left(\frac{\kappa}{0.57 \mbox{\ km s$^{-1}$ pc$^{-1}$}}\right) 
           \mbox{M$_\odot$ pc$^{\rm -2}$},
\end{equation}
and an azimuthally averaged gas surface density 
$\Sigma_{\rm gas} \sim 1.0 \times 10^3$ M$_\odot$ pc$^{\rm -2}$.
Then we obtain the $Q$ value of 0.6, indicating that the molecular gas 
in the ring is gravitationally unstable.
Therefore dense molecular clouds could be formed 
via gravitational instability of the molecular gas in the circumnuclear
star forming ring of NGC 6951. 
It is then quite interesting to evaluate $Q$ values 
in the outer regions of the ring where less active star formation occurs 
despite of strong CO emission, but it appears to be no longer valid to apply 
the Toomre's $Q$ stability analysis in these regions 
because the gas morphology is far from an axisymmetric disk 
and the deviations of the gas orbits 
from circular motion are also significant there. 
We speculate that the molecular gas in the CO peaks may not be gravitationally 
unstable, probably due to the very large velocity dispersion there.

Next, we evaluated the time scale of the gravitational instability.
The growth rate of the gravitational instability
can be estimated as the Jeans time scale (e.g. \cite{lar87}),
\begin{equation}
t_{\rm J} = \frac{\sigma_v}{\pi G \Sigma_{\rm gas}} 
       \simeq 1.1 \times 10^{\rm 6} \times 
       \left(\frac{\sigma_{v}}{15 \mbox{\ km s$^{\rm -1}$}}\right)
       \left(\frac{\Sigma_{\rm gas}}{1.0 \times 10^3 \mbox{\ M$_\odot$ pc$^{\rm -2}$}} 
             \right)^{\rm -1}
       \mbox{yrs,}
\end{equation}
for the averaged $\sigma_v$ and $\Sigma_{\rm gas}$ at the radius of the ring.
The rotation velocity at this radius is $\sim$ 160 km s$^{\rm -1}$ from 
Fig. \ref{fig13}, and the dynamical time scale at this radius 
$t_{\rm dyn} \sim 1.8 \times 10^{\rm 7}$ yrs.
The time scale of the instability $t_{\rm J}$ does not conflict 
with the observed offset between the CO and HCN peaks,
which is $t_{\rm offset} \sim 1/6 \,\, t_{\rm dyn} = 3.1 \times 10^{\rm 6}$ yrs 
at the radius of the ring.

In consequence, the gravitational instability scenario could nominally explain 
the features (i) and (ii).
We therefore suggest that {\it gravitational instabilities of molecular gas are 
responsible for the dense molecular gas formation and successive star formation
in the circumnuclear region of NGC 6951}.
This is sketched in Fig. \ref{fig17};
the molecular gas along the dust lanes and successive CO arms are too turbulent 
to form dense molecular clouds,
although the largest CO concentrations appear there. 
Some of the molecular gas escaped from the $x_1/x_2$ shocked
regions is driven into the $x_2$ orbits, and become unstable ($Q \leq 1$) 
because of the smaller velocity dispersion on the $x_2$ orbit ring
\footnote{
We note that Peng et al.\ (1996) have proposed that intense star formation 
occurs mostly in clouds on the $x_2$ orbits in the central region of 
nuclear starburst galaxy NGC 253,
which may be similar to the case in NGC 6951.
}.
The dense molecular cores are formed via gravitational instabilities.
The instabilities grow at a time scale of 
$t_{\rm J} \sim 1 \times 10^{\rm 6}$ yrs, which does not conflict 
with the observed offset between the CO peaks and HCN peaks, 
$t_{\rm offset} \sim 3 \times 10^{\rm 6}$ yrs.
Massive star formation then will occur in these dense molecular clouds,
showing a good coincidence between 
HCN emission and 6 cm radio continuum (or H$\alpha$) emission. 

\placefigure{fig17}

The degree of gravitational stabilities seems to rule the formation
of dense molecular clouds and successive star formation in other galaxies;
in the nuclear starburst galaxy NGC 3504, the molecular gas is 
gravitationally unstable over the rigid rotating part of the rotation curve, 
where the starburst occurs (\cite{kcy93}), and the dense molecular gas 
is concentrated in the same region as the starburst (\cite{koh97}).
On the other hand, the molecular gas in the central a few hundred parsec
region of NGC 4736 is found to be gravitationally {\it stable}
(\cite{shi98}; \cite{sak98}; \cite{koh98}), and we have found 
a remarkable decrease of HCN emission 
toward the central region of this galaxy (\cite{koh97}), 
despite the fact that the CO emission is concentrated there (\cite{sak98}). 
These results suggest again the presence of a strong connection 
between the degree of gravitational stabilities of the gas 
and the dense molecular gas formation, as in the case of NGC 6951.
Note that the nucleus of NGC 4736 is known as a ``post-starburst'' 
(\cite{pri77}; \cite{rlw88}; \cite{wlr88}; \cite{tan96}),
where its current star formation rate is significantly depressed.
Therefore the observed $R_{\rm HCN/CO}$ value of $\leq 0.04$
(2 $\sigma$ upper limit)
in the post-starburst region of NGC 4736 (\cite{koh97}),
which is much lower than starbursts 
($R_{\rm HCN/CO} \sim 0.1 - 0.2$; see Section 3.2),
also supports the existence of a tight connection 
between dense molecular gas and massive star formation
(Solomon et al.\ 1992; \cite{jac96}; \cite{pag97}).
There are other objects where the gravitational instabilities may govern 
the star formation, such as NGC 4102 (\cite{jak96}), NGC 4314
(\cite{ken97}), NGC 4414 (\cite{sak96}; \cite{bbb97})
and NGC 7331 (\cite{sat97}), although no HCN maps are available
for these galaxies yet.
In a much larger ($\geq$ a few kpc) scale, Kennicutt (1989) has
successfully explained the threshold gas surface density 
for the star formation in the disk of galaxies 
via gravitational instabilities of the gas.

Although Kennicutt (1989) assumed a constant gas velocity dispersion 
in his work, variation of gas velocity dispersion may be intimately
related to the gravitational instability of the gas 
as pointed out by Sakamoto (1996).
Contini et al. (1997) claimed that the older starbursts
have larger CO velocity dispersion in a kilo-parsec scale.
At the GMC scale, the GMCs which have larger velocity dispersion possess 
less active star formation (\cite{ias97}). 
Perhaps these results also imply the importance of the gas velocity dispersion 
on star formations at various spatial scales from GMCs to kilo-parsec scale 
galactic disks.

\subsection{Dense Molecular Gas toward the Seyfert Nucleus of NGC 6951}

The $R_{\rm HCN/CO}$ averaged over the central $r < 120$ pc region
was 0.086 toward the Seyfert nucleus in NGC 6951 (Fig. \ref{fig8}).
This $R_{\rm HCN/CO}$ value is in contrast to NGC 1068 and M51,
where very high $R_{\rm HCN/CO}$ values of 0.5 were reported
(\cite{jac93}; \cite{tac94}; \cite{hab95}; \cite{koh96}; \cite{mat98}).
What makes the difference of $R_{\rm HCN/CO}$ in these Seyfert nuclei?

A difference of CO and/or HCN fractional abundance may be a possible 
explanation because the strong X-ray radiation from the active nucleus 
has been claimed to cause the extremely high $R_{\rm HCN/CO}$ reported 
in NGC 1068 (\cite{lad96}; \cite{sag96}).
Therefore it is useful to compare the X-ray luminosity in these Seyfert 
galaxies, however, the intrinsic X-ray luminosity is quite difficult 
to measure due to the heavy absorption along the line of the sight 
in these type-2 AGN even in hard X-ray band. 

On the other hand, very high $R_{\rm HCN/CO}$ can be explained
without such abundance effect.
Matsushita et al.\ (1998) found that the molecular gas toward the center of M51 
is characterized by an extreme physical condition, that is,
a high temperature ($T_{\rm kin}$ $\gtrsim$ 100 K) and
high density ($n_{\rm H_2}$ $\sim$ 10$^{\rm 5 \pm 1}$ cm$^{\rm -3}$).
It should be noted that this result does scarcely depend on the adopted 
fractional HCN abundance.
Very high temperature and high density molecular gas has suggested
even in the central region of NGC 1068 (\cite{tac94}; \cite{hab95}).
Meanwhile, the observed $R_{\rm HCN/CO}$ of 0.086 in NGC 6951 
is similar to the value in the central $\sim$ 600 pc region of the Milky Way, 
0.08 (\cite{jac96}; \cite{hab97a})
where considerable amount of low density 
($n_{\rm H_2} \sim 10^2$ cm$^{\rm -3}$) molecular gas 
exist as well as high density molecular gas (e.g. \cite{oka98}). 
Therefore we conclude that the rather normal $R_{\rm HCN/CO}$ in NGC 6951 
may be attributed to the different physical condition of the molecular gas 
in NGC 6951 from those in NGC 1068 and M51.

We speculate that the difference of physical conditions of the molecular gas 
might be related to the existence of large scale (kpc scale) nuclear jets 
and ionized gas outflow;
both NGC 1068 and M51 possess a kpc scale jet, whereas NGC 6951 does not.
The projected length of radio jet in NGC 1068 and M51 is $\sim$ 1.3 kpc 
(\cite{wau87}; \cite{gal96}) and $\sim$ 1 kpc (\cite{for85}; \cite{cav92}), 
respectively. 
On the contrary, the central peak seen in a 20 cm radio continuum map 
(\cite{vil90}) is not resolved in the $\sim$ 1$''$ beam, imposing 
an upper limit of $\sim$ 120 pc for the length of the radio jet if it exists
\footnote{The 6 cm radio continuum map in Fig. \ref{fig5} (\cite{sai94})
shows a bridge like extension from the nucleus, however, this is the effect
of the elongated beam size. In fact, an higher resolution 6 cm image clearly 
shows an unresolved nuclear source (Saikia, private communication)}. 
Although these lengths are projected ones, a large scale jet in a pole-on 
configuration can not be likely in NGC 6951 because this is a type-2 AGN.
Therefore the presence of dense 
($n_{\rm H_2}$ $\sim$ 10$^{\rm 5}$ cm$^{\rm -3}$) molecular material 
toward the nucleus may be a result or cause of the nuclear jet/outflow. 
Indeed, an interaction between nuclear jet/outflow and molecular gas has been 
suggested in NGC 1068 in order to explain the beam filling factor of the HCN 
emission which is comparable with that of the CO (\cite{tac94}).
That would be also the case in M51; the line width of the HCN(1$-$0) emission 
toward the nucleus of M51 is wider than that of the CO(1$-$0) emission 
by a few tens km s$^{\rm -1}$ and this may also imply an interaction 
between dense molecular gas and the nuclear jet/outflow (\cite{koh96}). 
The asymmetric line profile from dense molecular gas is clearly seen 
even in the CO(2$-$1) emission (\cite{sco98}).
Alternatively, these dense molecular gas plays a role for the confinement of 
the jet/outflow (e.g. \cite{ias88}; \cite{bli94}; \cite{gal96}; \cite{sco98}).

\section{Conclusions}

We present high resolution (3$''$ -- 5$''$) observations of
CO(J=1$-$0) and HCN(J=1$-$0) emission from the circumnuclear
star forming ring in the barred spiral galaxy NGC 6951, a host of
a type-2 Seyfert nucleus, using the Nobeyama Millimeter Array 
and the Nobeyama 45 m telescope.
The results of our new observations and conclusions drawn from them
are summarized as follows:

\begin{enumerate}

\item We find that the distribution of the HCN emission is different 
from that of CO in the circumnuclear region of NGC 6951; 
it is confirmed that CO emission is dominated by a ``twin peaks'' morphology 
with two spiral arms, which are connected to the dust lanes, 
as reported by Kenney et al. (1992).
On the other hand, although the HCN emission also shows a twin peaks 
morphology, the HCN peaks are spatially shifted downstream compared 
with the CO peaks.

\item Most of the HCN emission is associated with the circumnuclear ring,
where vigorous star formation occurs.
In other words, the HCN emission spatially correlates better
with the massive star forming regions
than that of CO emission in the circumnuclear region of NGC 6951.

\item The HCN to CO integrated intensity ratio in the brightness temperature 
scale, $R_{\rm HCN/CO}$, is enhanced in the star forming ring.
The peak value of the $R_{\rm HCN/CO}$ is about 0.16 -- 0.18,
which is comparable to the $R_{\rm HCN/CO}$ in the starburst regions of 
NGC 253 and M82.

\item The formation mechanism of dense molecular gas has been investigated. 
No significant enhancement of $R_{\rm HCN/CO}$ is observed at the CO peaks,
which are interpreted as $x_1/x_2$ orbit crowding regions.
This suggests that the shocks along the orbit crowding do not promote
the formation of the dense molecular gas effectively
but enhance the presence of low density GMCs in NGC 6951.
Instead, gravitational instability can account for the dense molecular gas
formation in the circumnuclear star forming ring because
Toomre's $Q$ value is below unity there.

\item $R_{\rm HCN/CO}$ is 0.086 averaged over the central $r < 120$ pc region
toward the nucleus of NGC 6951, which shows a type-2 Seyfert activity.
This is a rather normal value compared with non-active galaxies
such as the Milky Way,
and quite different from other type-2 Seyfert galaxies NGC 1068 and M51
where extremely high $R_{\rm HCN/CO}$ value, $\sim 0.5$, are reported.
The variety of $R_{\rm HCN/CO}$ value in these Seyfert nuclei
would be attributed with the different physical conditions 
of the molecular gas around the active nuclei.
The presence of large scale nuclear jets/outflows might be related to these
difference of molecular gas properties.

\end{enumerate}

\acknowledgments

We would like to thank the referee, Dr. J. D. P. Kenney, 
for the critical reading of the manuscript and invaluable suggestions 
which greatly improved this work.
We are grateful to Dr. H. Wozniak and Dr. P. Martine 
for sending us their H$\alpha$ image of NGC 6951, 
Dr. A. J. Barth for his HST V-band image,
and Dr. D. J. Saikia for providing us with his not yet published 
new radio continuum map.
We thank S. Takakuwa for providing us with his LVG code.
We thank T. Kamazaki for assistance with making the figures.
We are deeply indebted to the NRO staff for the operation of the telescopes 
and the continuous efforts in improving the performance of the instruments. 
K. K. was financially supported by the Japan Society for the Promotion 
of Science.

\clearpage

\figcaption{An optical image from the Digitized Sky Survey 
and H$\alpha$ + N[II] contour map (white contours; Wozniak et al.\ 1995)
in the central 3$'$ $\times$ 3$'$ (21 kpc $\times$ 21 kpc
at $D$ = 24.1 Mpc) field of NGC 6951. 
Contour levels of the H$\alpha$ + N[II] image are
1.0, 2.5, 5.0, 10.0, 15.0, 20.0, and 30.0 $\times$ 10$^{\rm -15}$
erg s$^{\rm -1}$ cm$^{\rm -2}$ arcsec$^{\rm -2}$.
The boundary of the H$\alpha$ + N[II] map is indicated by white lines.
The mapped area with the NMA is also denoted as white dashed circles;
the smaller and larger circles represent the NMA primary beam size
at CO(1$-$0) observations (65$''$ HPBW) and HCN(1$-$0) observations
(84$''$), respectively.
It is evident that most of the star formation is confined within the 
circumnuclear ring.
\label{fig1}}

\figcaption{Velocity channel maps of CO(1$-$0) and HCN(1$-$0) emission
in the central 30$''$ $\times$ 30$''$ 
(3.5 kpc $\times$ 3.5 kpc at $D$ = 24.1 Mpc) region of NGC 6951.
The cross shows the position of the nucleus defined 
as the central peak of 6 cm radio continuum (Saikia et al.\ 1994).
The attenuation due to the primary beam pattern is not corrected in these maps.
(a) Channel maps of CO emission.
The velocity width of each channels is 9.79 km s$^{\rm -1}$
and the central velocities ($V_{\rm LSR}$ in km s$^{\rm -1}$) are labeled.
The beam size is $3\farcs9 \times 3\farcs1$ with $P.A.$ of $0^\circ$.
Contour levels are -4, -2, 2, $\cdots$, 10,  and 12 $\sigma$, 
where 1 $\sigma$ = 34 mJy beam$^{\rm -1}$ or 260 mK in $T_{\rm b}$.
Negative contours are dashed.
(b) Channel maps of HCN emission.
Velocity width of each channel is 38.2 km s$^{\rm -1}$
nd the central velocities ($V_{\rm LSR}$ in km s$^{\rm -1}$) are labeled.
The beam size is $4\farcs7 \times 4\farcs5$, $P.A.$ $165^\circ$.
Contour levels are -3, -1.5, 1.5, 3, 4.5, and 6 $\sigma$, 
where 1 $\sigma$ = 5.1 mJy beam$^{\rm -1}$ or 37 mK in $T_{\rm b}$.
\label{fig2}}

\figcaption{Integrated intensity maps and line profile maps of 
the CO(1$-$0) and HCN(1$-$0) emission, derived from the NMA data cubes, 
in the central 20$''$ $\times$ 25$''$ 
(2.3 kpc $\times$ 2.9 kpc) region of NGC 6951.
The line profiles were measured at 5 $\times$ 7 points 
on the 3$''$ spacing $\alpha$ -- $\delta$ grid.
The grid positions are indicated by crosses in the integrated intensity maps.
The central grid position corresponds to the nucleus 
defined as the 6 cm radio continuum peak (Saikia et al.\ 1994).
The attenuation due to the primary beam pattern is corrected in these maps. 
(a) Integrated intensity map of CO(1$-$0) emission
over a velocity range from $V_{\rm LSR}$ = 1223 to 1564 km s$^{\rm -1}$.
The beam size is $3\farcs9 \times 3\farcs1$ with $P.A.$ of $0^\circ$.
The contour levels of the CO map are 2, 4, 6, $\cdots$, and 26 $\sigma$
where 1 $\sigma$ is 1.9 Jy beam$^{\rm -1}$ km s$^{\rm -1}$ or 15 K km s$^{\rm -1}$,
corresponding to a face-on molecular gas surface density 
$\Sigma_{\rm gas} = 1.36 \times \Sigma_{\rm H_2} = 73$ $M_\odot$ pc$^{-2}$.
(b) Integrated intensity map of HCN(1$-$0) emission
over a velocity range from $V_{\rm LSR}$ = 1225 to 1567 km s$^{\rm -1}$
The beam size is $4\farcs7 \times 4\farcs5$, $P.A.$ 165$^\circ$.
The contour levels are 2, 3, 4, $\cdots$, and 9 $\sigma$, 
where 1 $\sigma$ = 0.58 Jy beam$^{\rm -1}$ km s$^{\rm -1}$  or 4.3 K km s$^{\rm -1}$.
(c) CO line profiles. The noise level is 34 mJy beam$^{\rm -1}$ 
or 260 mK in $T_{\rm b}$ at the central position. 
The peak brightness temperature is $\sim$ 3.5 K.
The width of a velocity bin is 9.79 km s$^{\rm -1}$.
(d) HCN line profiles. The noise level is 5.1 mJy beam$^{\rm -1}$ 
or 38 mK in $T_{\rm b}$ at the central position. 
The peak brightness temperature is $\sim$ 0.38 K. 
The width of a velocity bin is 19.1 km s$^{\rm -1}$.
\label{fig3}}

\figcaption{CO(1$-$0) and HCN(1$-$0) spectra toward the nucleus of NGC 6951
taken with the NRO 45 m telescope (dashed line),
together with spectra from the NMA observations (solid line).
The velocity resolution of the CO spectrum is 10 km s$^{\rm -1}$ and 
HCN is 20 km s$^{\rm -1}$ for the 45 m spectra.
Beam size (HPBW) at the frequencies of CO(1$-$0) and HCN(1$-$0) are 
15$''$ and 19$''$, respectively.
The spectra from the NMA data are made after convolving the data cube 
to the same beam size as the 45 m beam.
\label{fig4}}

\figcaption{Comparison of the CO(1$-$0) and HCN(1$-$0) distributions (thin contours)
with the H$\alpha$ + [NII] (greyscale, Wozniak et al. 1995) 
and 6 cm radio continuum images (thick contours, Saikia et al.\ 1994).
The seeing size of the H$\alpha$ + [NII] is about 1$\farcs$5.
The greyscale ranges from 1.30 $\times$ 10$^{\rm -15}$ to 3.07 $\times$ 10$^{\rm -14}$
ergs s$^{\rm -1}$ cm$^{\rm -2}$ arcsec$^{\rm -2}$.
The beam size of 6 cm radio map is 2$\farcs$57 $\times$ 1$\farcs$44 with 
a $P.A.$ of 80$^\circ$, and contour levels are 3, 6, 9, $\cdots$, and 18 $\sigma$,
where 1 $\sigma$ = 60 $\mu$Jy beam$^{\rm -1}$.
The traces of dust lanes seen in a $B-I$ color image (Wozniak et al.\ 1995)
are denoted as dashed lines. 
The ellipses show the synthesized beams of CO and HCN observations. 
The direction of the galactic rotation is indicated by an arrow.
It is evident that the distribution of HCN emission is different from that of CO;
the HCN peaks are spatially shifted downstream with respect to the CO peaks,
and the most of HCN emission are associated with the circumnuclear star forming ring
traced by H$\alpha$ and radio continuum emission.
\label{fig5}}

\figcaption{Curves of constant $R_{\rm HCN/CO}$ (solid lines) 
and brightness temperature of the HCN(1$-$0) emission (dashed lines) 
as functions of kinetic temperature and density of molecular gas. 
CO fractional abundance per unit velocity gradient $Z$(CO)/($dv/dr$) 
was set to $10^{-4}$, where $Z$(CO) is defined as [CO]/[H$_{2}$],
and the unit of $dv/dr$ is km s$^{-1}$ pc$^{-1}$. The fractional abundance ratio
of HCN and CO molecules were assumed to be $Z$(HCN)/$Z$(CO) = 1/4000 
(Irvine, Goldsmith, \& Hjalmarson 1987).
\label{fig6}
}

\figcaption{(a) Map of the HCN(1$-$0)/CO(1$-$0) integrated intensity ratio in
brightness temperature scale, $R_{\rm HCN/CO}$.
Contour levels are 0.08, 0.1, 0.12, $\cdots$, and 0.18.
Greyscale ranges from 0.08 to 0.2.
The traces of dust lanes seen in a $B-I$ color image (Wozniak et al.\ 1995)
are denoted as dashed lines. 
(b) 6 cm radio continuum map (Saikia et al.\ 1994) superposed 
on the greyscale image of the $R_{\rm HCN/CO}$ distribution.
It is clear that the $R_{\rm HCN/CO}$ is enhanced at the massive star forming 
regions traced by the radio continuum emission. 
\label{fig7}}

\figcaption{Radial profile of the $R_{\rm HCN/CO}$ in the central region of NGC 6951.
The ratio was averaged over the successive annulli with 1$''$ = 117 pc width.
The location of the circumnuclear star forming ring 
traced by H$\alpha$ emission is also indicated. 
\label{fig8}}

\figcaption{Definition of the polar coordinate for Fig. 10. 
$\phi$ = 0$^\circ$ corresponds to the major axis of the galaxy.
\label{fig9}}

\figcaption{CO, HCN, and $R_{\rm HCN/CO}$ distributions on $R - \phi$ coordinate 
denoted as Fig. 9.
Contour levels of CO and HCN maps are same as Fig. 3.
Contour levels of $R_{\rm HCN/CO}$ are 0.1, 0.12, 0.14, 0.16, and 0.18.
The peak positions of CO and $R_{\rm HCN/CO}$ are indicated as dashed lines.
We see the spatial offset of HCN peaks with respect to the CO peaks,
which is also evident in $R_{\rm HCN/CO}$ map.
The coincidence of the dense molecular gas and massive star forming regions
traced by the 6 cm radio continuum is also clearly shown.
\label{fig10}}

\figcaption{Intensity-weighted mean velocity map of the CO emission
in the central region of NGC 6951, superposed on the grey-scale image
of the velocity integrated intensity map of the CO emission.
The contour interval is 20 km s$^{\rm -1}$, and contours are
labeled by the value of $V_{\rm LSR}$.
The contour which represents the systemic velocity, 1428 km s$^{\rm -1}$,
is displayed by a thick contour.
Note that the ``S-shape'' contours near the systemic velocity, indicating
the presence of non-circular motions there.
\label{fig11}}

\figcaption{Position - velocity map ($pv$ map) of the CO emission
in NGC 6951 along the major axis ($P.A.$ = 130$^\circ$). 
Contour levels are -3, -1.5, 1.5, $\cdots$, and 12 $\sigma$,
where 1 $\sigma$ = 34 mJy beam$^{\rm -1}$ or 260 mK in $T_{\rm b}$.
Negative contours are dashed.
The cross indicates the kinematical center
($\delta$(B1950) = +65$^\circ$ 55$\arcmin$ 46$\farcs$0,
$V_{\rm sys}$ = 1428 km s$^{\rm -1}$). 
\label{fig12}}

\figcaption{(a) Circular rotation curve of NGC 6951 derived from the
intensity-weighted mean velocity map of the CO emission. 
This is derived from the intensity-weighted mean velocity mapof the CO emission
within $\pm 5^\circ$ of the major axis.
Inclination (42$^\circ$) is corrected.
(b) Angular velocities $\Omega$, epicyclic frequency $\kappa$, and 
$\Omega$ - $\kappa$/2 as a function of galactocentric radius, 
derived from the CO circular rotation curve (a).
The $\kappa$ is divided by 2 to facilitate comparison.
The hofizontal line (dashed line) indicates an estimation 
of the bar pattern speed, 50 km s$^{\rm -1}$ kpc$^{\rm -1}$.
Horizontal arrows show the loci of the ring and CO arms.
\label{fig13}}

\figcaption{CO velocity residuals produced by subtracting the rotation curve 
in Fig. 12(a)
from the observed mean velocity field in Fig. 10. 
Residuals are superposed on the integrated CO intensity greyscale map.
The velocity interval is 15 km s$^{\rm -1}$.
Dust lanes are indicated by straight dashed lines.
The largest deviations are seen on the CO arms which are connected to the dust lanes.
\label{fig14}}

\figcaption{Intensity-weighted velocity dispersion map of the CO emission.
The contour interval is 10 km s$^{\rm -1}$.
This map is computed by calculating the second order moment of
intensity at each point from the CO cube.
Note that this map contains both the intrinsic velocity dispersion of gas
and the gradient of rotation velocity within the observed beam.
\label{fig15}}

\figcaption{Comparison of $R_{\rm HCN/CO}$ ratio distribution (greyscale) 
with contour maps of (a) the integrated CO intensity, 
and (b) the 2nd moment map of the CO cube.
The greyscale ranges from 0.13 to 0.18.
Contour intervals of (a) and (b) are same as Fig. 3(a) and Fig. 15.
It is evident that the $R_{\rm HCN/CO}$ ratio is not enhanced at the CO arms, 
which is thought to be a shocked region associated with the $x_1/x_2$ orbit crowding.
There also exist a possible anti-correlation between the $R_{\rm HCN/CO}$ 
and the gas velocity dispersion. 
\label{fig16}}

\figcaption{Schematic view of the CO and HCN morphology, and the gas motion 
in the central region of NGC 6951. 
The molecular gas along the dust lanes and successive CO arms are too turbulent
to form dense molecular clouds,
although the largest CO concentrations appear there.
Some of the molecular gas escaped from the $x_1/x_2$ shocked
regions is driven into the $x_2$ orbits, and become unstable ($Q < 1$)
because of the smaller velocity dispersion on the $x_2$ orbit ring,
and dense molecular cores are formed via gravitational instabilities.
Massive star formation then will occur in these dense molecular cores,
showing a good coincidence between HCN emission and 6 cm radio continuum 
(or H$\alpha$) emission.
\label{fig17}}

\clearpage
 
\begin{deluxetable}{llc}
\tablewidth{0pt}
\tablecaption{Properties of NGC 6951. \label{tbl1}}
\tablehead{
\colhead{Parameter}  & \colhead{Value}  &  \colhead{Ref.}
}
\startdata
Morphology                         & SAB(rs)bc                      & RC3 \nl
                                   & SB/SBb(rs)I.3                  & RSA \nl
Nuclear activity                   & Type 2 Seyfert                 & (1) \nl
Position of nucleus                &                                & (2) \nl
\ \ \ $\alpha$ (B1950)             & 20$^{\rm h}$36$^{\rm m}$36$\fs$59 & \nl
\ \ \ $\delta$ (B1950)             & +65$^\circ$55$\arcmin$46$\farcs$0 & \nl
$D_{25}$ $\times$ $d_{25}$         & 3$\farcm$9 $\times$ 3$\farcm$2 & RC3 \nl
Position angle                     & 130$^\circ$                    & (3) \nl
Inclination angle                  & 42$^\circ$                     & (4) \nl
Adopted Distance                   & 24.1 Mpc                       & (5) \nl
Linear scale                       & 117 pc arcsec$^{\rm -1}$       &     \nl
$S_{{\rm 60} \mu{\rm m}}$          & 13.21 $\pm$ 0.528 Jy           & (6) \nl
$S_{{\rm 100} \mu{\rm m}}$         & 37.47 $\pm$ 1.498 Jy           & (6) \nl
$L_{\rm FIR}$\tablenotemark{a}     & 1.6 $\times$ 10$^{\rm 10}$ $L_\odot$  & $\cdots$ \nl
$L_{{\rm H}\alpha}$ (ring)\tablenotemark{b}
                                   & 4.2 $\times$ 10$^{\rm 41}$ ergs s$^{\rm -1}$ & $\cdots$ \nl
SFR (ring)\tablenotemark{c}        & 3.8 $M_\odot$ yrs$^{\rm -1}$   & $\cdots$  \nl
\tablenotetext{a}{$L_{\rm FIR}$ was calculated as
$3.75 \times 10^5 \times (D/\mbox{Mpc})^2 
(2.58 S_{{\rm 60} \mu{\rm m}} + S_{{\rm 100} \mu{\rm m}})$
in $L_\odot$.
}
\tablenotetext{b}{Measured from the central 13$''$ $\times$ 13$''$ aperture
in an H$\alpha$ + [NII] image (Wozniak et al.\ 1995).
Contributions from the central 3$''$ $\times$ 3$''$ aperture 
(Seyfert nucleus) was subtracted.
Internal extinction was corrected assuming $A_{\rm v}$ = 3.4 mag (Barth et al.\ 1995).
$EW$(H$\alpha$ + [NII]) = 1.33 $\times$ $EW$(H$\alpha$) was adopted (Kennicutt 1983).
}
\tablenotetext{c}{SFR was derived from
$L_{{\rm H}\alpha}$/(1.12 $\times$ 10$^{\rm 41}$ ergs s$^{\rm -1}$)
in $M_\odot$ yrs$^{\rm -1}$ (Kennicutt 1983).
}
\tablerefs{
(1) Ho et al.\ 1995; (2) Saikia et al.\ 1994; (3) This work;
(4) M\'arquez \& Moles 1993; (5) Tully 1988; (6) Ho et al.\ 1997
}
\enddata
\end{deluxetable}

\clearpage
 
\begin{deluxetable}{lll}
\tablewidth{0pt}
\tablecaption{NMA observations and results. \label{tbl2}}
\tablehead{ \colhead{Parameter}  & \colhead{Value} & \colhead{} }
\startdata
Observing date        & \multicolumn{2}{l}{Nov. 1995 - Feb. 1996} \nl
Number of antennas    & \multicolumn{2}{l}{6}                     \nl
Array configuration   & \multicolumn{2}{l}{AB, C, \& D}           \nl
Phase center          & & \nl
\ \ \ $\alpha$(B1950)& \multicolumn{2}{l}{20$^{\rm h}$36$^{\rm m}$36$\fs$59} \nl
\ \ \ $\delta$(B1950)& \multicolumn{2}{l}{+65$^\circ$55$\arcmin$46$\farcs$0} \nl
Visibility calibrator & \multicolumn{2}{l}{3C418} \nl
Band width            & \multicolumn{2}{l}{319.815680 MHz} \nl
\cutinhead{Line-specific parameters}
Line and transition   & CO(J=1$-$0)   & HCN(J=1$-$0) \nl
Rest frequency        & 115.271204 GHz  & 88.6316024\tablenotemark{a}\ \ GHz \nl
Observed frequency    & 114.725 GHz & 88.210 GHz \nl
Projected baseline    & 3.8 -- 73 k$\lambda$ & 2.9 -- 102 k$\lambda$ \nl
Field of view         & 65$''$ (7.6 kpc) & 84$''$ (9.8 kpc) \nl
Velocity coverage     & 836 km s$^{\rm -1}$ & 1088 km s$^{\rm -1}$ \nl
Velocity resolution   & 9.79 km s$^{\rm -1}$  & 38.2 km s$^{\rm -1}$ \nl
Synthesized beam      & 3$\farcs$9 $\times$ 3$\farcs$1, $P.A.$ 0$^\circ$
                      & 4$\farcs$7 $\times$ 4$\farcs$5, $P.A.$ 165$^\circ$ \nl
                      & (460 $\times$ 360 pc) & (550 $\times$ 530 pc) \nl
Equivalent $T_{\rm b}$ for 1 Jy beam$^{\rm -1}$
                      & 7.7 K       &  7.4 K  \nl
r.m.s. noise in channel map
                      & 34 mJy beam$^{\rm -1}$ & 5.1 mJy beam$^{\rm -1}$ \nl
                      & = 260 mK in $T_{\rm b}$ & = 38 mK in $T_{\rm b}$ \nl
NMA flux within F.O.V. & 334 $\pm$ 12 Jy km s$^{\rm -1}$ & 18.7 $\pm$ 1.9 Jy km s$^{\rm -1}$ \nl
FCRAO flux             & 350 $\pm$ 41 Jy km s$^{\rm -1}$ & $\cdots$       \nl
S(NMA)/S(FCRAO)        & 0.95                           & $\cdots$       \nl
Luminosity $L'$        & 4.8 $\times$ 10$^{\rm 8}$ K km s$^{\rm -1}$ pc$^{\rm 2}$ &
                         4.5 $\times$ 10$^{\rm 7}$ K km s$^{\rm -1}$ pc$^{\rm 2}$   \nl
Luminosity ratio $L_{\rm FIR}/L'$\tablenotemark{b} & 38 & 390 \nl
\enddata
\tablenotetext{a}{weighted mean of three hyper-fine transitions}
\tablenotetext{b}{$L_{\rm FIR}/L'_{\rm CO}$ and $L_{\rm FIR}/L'_{\rm HCN}$, 
respectively. $L_{\rm FIR}$ is in Table 1.
}
\tablecomments{The uncertainties of fluxes are statistical errors only.}
\end{deluxetable}

\clearpage
 
\begin{deluxetable}{lll}
\tablewidth{0pt}
\tablecaption{45 m observations and results. \label{tbl3}}
\tablehead{ \colhead{Parameter}  & \colhead{CO(J=1$-$0)} & \colhead{HCN(J=1$-$0)} }
\startdata
Beam size (HPBW) & 15\arcsec & 19\arcsec \nl
Main beam efficiency $\eta_{\rm MB}$ & $0.45 \pm 0.03$ & $0.50 \pm 0.03$ \nl
Peak Temperature $T_{\rm MB}$  & 380 $\pm$ 21 mK & 38 $\pm$ 4.7 mK \nl
Velocity width (FWZI) & 380.2 $\pm$ 10 km s$^{\rm -1}$ 
                      & 364.2 $\pm$ 20 km s$^{\rm -1}$ \nl
Integrated intensity $\int T_{\rm MB}(v)dv$
                            & 71 $\pm$ 8.0 K km s$^{\rm -1}$ 
                            & 6.5 $\pm$ 1.7 K km s$^{\rm -1}$ \nl
45 m flux\tablenotemark{a}  & 172 $\pm$ 19 Jy km s$^{\rm -1}$ 
                            & 15.0 $\pm$ 3.9 Jy km s$^{\rm -1}$ \nl
NMA flux in same beam\tablenotemark{b} & 197 Jy km s$^{\rm -1}$  
                                       & 14.1 Jy km s$^{\rm -1}$ \nl
S(NMA)/S(45m)         & 1.1  & 0.94 \nl
Luminosity $L'$  & 2.4 $\times$ 10$^{\rm 8}$ K km s$^{\rm -1}$ pc$^{\rm 2}$
                 & 3.6 $\times$ 10$^{\rm 7}$ K km s$^{\rm -1}$ pc$^{\rm 2}$ \nl
\enddata
\tablenotetext{a}{The CO and HCN fluxes calculated from the integrated intensities.
The point source sensitivities of 2.4 Jy K$^{\rm -1}$ and 2.3 Jy K$^{\rm -1}$
are assumed for CO and HCN observations, respectively.}
\tablenotetext{b}{The CO and HCN fluxes measured from the convolved cubes
as the same beam size of 45 m observations.}
\tablecomments{The uncertainties of fluxes are statistical errors only.}
\end{deluxetable}
 
\clearpage
 
\begin{deluxetable}{lll}
\tablewidth{0pt}
\tablecaption{Kinematical parameters derived from the CO velocity field\tablenotemark{a}. \label{tbl4}}
\tablehead{ \colhead{Parameter}  & \colhead{Value} }
\startdata
Kinematical center     & \nl
\ \ \ $\alpha$ (B1950) & 20$^{\rm h}$36$^{\rm m}$36$\fs$66 $\pm$ 0$\fs$16 & \nl
\ \ \ $\delta$ (B1950) & +65$^\circ$55$'$46$\farcs$3 $\pm$ 1$''$ & \nl
Position angle      & 130$^\circ$ $\pm$ 4$^\circ$ \nl
Systemic velocity\tablenotemark{b}   & 1428 km s$^{\rm -1}$ $\pm$ 2 km s$^{\rm -1}$ \nl
\enddata
\tablenotetext{a}{fitting was made within the $1'' < r < 5''$ region.}
\tablenotetext{b}{LSR velocity. Subtract 14.4 km s$^{-1}$ to obtain heliocentric velocities.}
\end{deluxetable}


\clearpage


\begin{thebibliography}{}

\bibitem[Aalto et al.\ 1995]{aal95} Aalto, S. Booth, R. S., Black, J. H.,
    \& Johansson, L. E. B., 1995, \aap, 300, 369

\bibitem[Athanassoula 1992]{ath92} Athanassoula, E.  1992, \mnras, 259, 345

\bibitem[Barth et al.\ 1995]{bar95} Barth, A. J., Ho, L. C., Filippenko, A. V.,
    \& Sargent W. L. W. 1995, \aj, 110, 1009

\bibitem[Benedict, Smith, \& Kenney 1996]{bsk96} Benedict, G. F., 
    Smith, B. J., \& Kenney, J. D. P.  1996, \aj, 111, 1861

\bibitem[B\"oker, F\"orster-Schreiber, \& Genzel 1997]{bfg97} B\"oker, T., 
    F\"orster-Schreiber, N. M., \& Genzel, R.  1997, \aj, 114, 1883

\bibitem[Blietz et al.\ 1994]{bli94} Blietz, M., Cameron, M., Drapatz, S., Genzel, R., Krabbe, A.,
    \& van der Werf, P.  1994, \apj, 421, 92

\bibitem[Boer \& Schulz 1993]{bas93} Boer, B., \& Schulz, H. 1993, 
    \aap, 227, 397

\bibitem[Bottinelli et al.\ 1984]{bot84} Bottinelli, L., 
    Gouguenheim, L., Paturel, G., \& Devaucouleurs, G.  1984,
    \aaps, 56, 381

\bibitem[Braine, Brouillet, \& Baudry 1997]{bbb97} Braine, J.,
    Brouillet, N., \& Baudry, A.  1997, \aap, 318, 19

\bibitem[Brouillet \& Schilke 1993]{bro93} Brouillet, N., \& Schilke, P. 
   1993, \aap, 277, 381

\bibitem[Buta \& Crocker 1993]{bac93} Buta, R., \& Crocker, D. A.
    1993, \aj, 105, 1344

\bibitem[Combes \& Gerin 1985]{cag85} Combes, F., \& Gerin, M.  
    1985, \aap, 150, 327

\bibitem[Contini et al.\ 1997]{con97} Contini, T., Wozniak, H.,
    Considere, S., \& Davoust, E.  1997, \aap, 318, 51

\bibitem[Crane \& van der Hulst 1992]{cav92} Crane, P. C.,
    \& van der Hulst, J. M.  1992, \aj, 103, 1146

\bibitem[RC3]{rc3} de Vaucouleurs, G., de Vaucouleurs, A., Corwin, H. G., 
    Buta, R. J., Paturel, G., \& Fouque, P.  1991, 
    Third Reference Catalogue of Bright Galaxies (New York: Springer-Verlag)

\bibitem[Downes et al.\ 1992]{dow92} Downes, D., Radford, S. J. E.,
    Guilloteau, S., Gu\'elin, M., Greve, A., \& Morris, D.  1992, \aap,
    262, 424

\bibitem[Downes et al.\ 1996]{dow96} Downes, D., Reynaud, D., Solomon, P. M.,
    Radford, S. J. E.  1996, \apj, 461, 186

\bibitem[Elmegreen 1994]{elm94} Elmegreen, B. G.,  1994, \apjl, 425, L73

\bibitem[Elmegreen et al.\ 1996]{elm96} Elmegreen, D. M., Elmegreen, B. G.,
    Chromey, F. R., Hasselbacher, D. A., \& Bissell, B. A.  1996, \aj,
    111, 1880

\bibitem[Ford et al.\ 1985]{for85} Ford, H. C., Crane, P. C.,
    Jacoby, G. H., Lawrie, D. G.  1985, \apj, 293, 132

\bibitem[Friedli et al.\ 1996]{fri96} Friedli, D., Wozniak, H., Rieke, M.,
    Martinet, L., \& Bratschi, P.\  1996, \aaps, 118, 461

\bibitem[Gao \& Solomon 1996]{gao96} Gao, Y., \& Solomon, P. M.  1996,
    in CO: Twenty-Five Years of Millimeter-Wave Spectroscopy,
    ed. Latter, W. B., Radford, S. J. E., Jewell, P. R., Mangum, J. G., \& Bally, J.,
    (Dordrecht: Kluwer), 419

\bibitem[Gallimore et al.\ 1996]{gal96} Gallimore, J. F., Baum, S. A., O'Dea, C. P.,
    \& Pedlar, A., 1996, \apj, 464, 198

\bibitem[Goldreich \& Kwan 1974]{gak74} Goldreich, P., \& Kwan, J.  
    1974, \apj, 189, 441

\bibitem[Golla, Allen, \& Kronberg 1996]{gol96} Golla, G., Allen, M. L., 
    \& Kronberg, P. P.  1996, \apj, 473, 244

\bibitem[Gonz\'alez-Delgado et al.\ 1997]{gon97} Gonz\'alez-Delgado, R. M.,
    P\'erez, E., Tadhunter, C., Vilchez, J. M., 
    \& Rodr\'{\i}gues-Espinosa, J. M.  1997, \apjs, 108, 155

\bibitem[G\"usten et al.\ 1993]{gus93} G\"usten, R., Serabyn, E., Kasemann, C.,
    Schinckel, A., Schneider, G., Schulz, A., \& Young, K.  1993, \apj, 402, 537

\bibitem[Henkel, Baan, \& Mauersberger 1991]{hen91} Henkel, C.,
    Baan, W. A., \& Mauersberger, R.  1991, \aapr, 3, 47

\bibitem[Helfer \& Blitz 1993]{hab93} Helfer, T., \& Blitz, L.  
    1993, \apj, 419, 86

\bibitem[Helfer \& Blitz 1995]{hab95} Helfer, T.,
    \& Blitz, L.  1995, \apj, 450, 90

\bibitem[Helfer \& Blitz 1997a]{hab97a} Helfer, T., \& Blitz, L.  
    1997, \apj, 478, 162

\bibitem[Helfer \& Blitz 1997b]{hab97b} Helfer, T., \& Blitz, L.  
    1997, \apj, 478, 233

\bibitem[Ho, Filippenko, \& Sargent 1995]{hfs95} Ho, L. C.,
    Filippenko, A. V., \& Sargent, W. L. W. 1995, \apjs, 98, 477

\bibitem[Ho, Filippenko, \& Sargent 1997]{hfs97} Ho, L. C.,
    Filippenko, A. V., \& Sargent, W. L. W. 1997, \apjs, 112, 315

\bibitem[Irvine, Goldsmith, \& Hjalmarson 1987]{irv87} Irvine, W. M.,
    Goldsmith, P. F., \& Hjalmarson, \AA.  1987,
    in Interstellar Processes, ed. D. J. Hollenbach \& H. A. Thronson, Jr.
    (Dordrecht: Reidel Publishing Company), 561

\bibitem[Irwin \& Seaquist 1988]{ias88} Irwin, J. A., \& Seaquist, E. R.  1988, \apj,
    335, 658

\bibitem[Ishizuki et al. 1990]{ish90} Ishizuki, S., Kawabe, R., Ishiguro, M.,
    Okumura, S. K., Morita, K. -I., Chikada, Y., Kasuga, T.  1990, Nature, 344, 224

\bibitem[Ikuta \& Sofue 1997]{ias97} Ikuta, C, \& Sofue, Y.  1997, \pasj, 49, 323

\bibitem[Jackson et al.\ 1993]{jac93} Jackson, J. M., 
    Paglione, T. D., Ishizuki, S., \& Rieu, N. Q.  1993,
    \apjl, 418, L13

\bibitem[Jackson et al.\ 1996]{jac96} Jackson, J. M.,
    Heyer, M. H., Paglione, T. A. D., \& Bolatto, A. D.  1996,
    \apjl, 456, L91 

\bibitem[Jogee \& Kenney 1996]{jak96} Jogee, S., \& Kenney, J. D. P.
    1996, in Barred Galaxies, ed. R. Buta, D. A. Crocker, \& B. G. Elmegreen
    (San Francisco: ASP), 230 

\bibitem[Kenney et al.\ 1992]{ken92} Kenney, J. D. P., Wilson, C. D.,
    Scoville, N. Z., Devereux, N. A., \& Young, J. Y. 1992, \apjl,
    395, L79

\bibitem[Kenney, Carlstrom, \& Young 1993]{kcy93} Kenney, J. D. P.,
    Carlstrom, J. E., \& Young, J. S.  1993, \apj, 418, 687

\bibitem[Kenney 1996]{ken96} Kenney, J. D. P.
    1996, in Barred Galaxies, ed. R. Buta, D. A. Crocker, \& B. G. Elmegreen
    (San Francisco: ASP), 150

\bibitem[Kenney 1997]{ken97} Kenney, J. D. P.  1997,
    in Starburst Activity in Galaxies, ed. J. Franco, R. Terlevich,
    \& A. Serrano (Mexico: Rev. Mex. Astron. Ap), 52 

\bibitem[Kennicutt 1983]{ken83} Kennicutt, R., C.  1983, \apj, 272, 54

\bibitem[Kennicutt 1989]{ken89} Kennicutt, R., C.  1989, \apj, 344, 685

\bibitem[Kohno et al.\ 1996]{koh96} Kohno, K., Kawabe, R., Tosaki, T.,
    \& Okumura, S. K.  1996, \apjl, 461, L29

\bibitem[Kohno et al.\ 1997]{koh97} Kohno, K., Kawabe, R., Sakamoto, K.,
    Ishizuki, S., \& Vila-Vilar\'o, B.  1997,
    in The Central Regions of Galaxy and Galaxies, ed. Y. Sofue
    (Dordrecht: Kluwer), in press

\bibitem[Kohno 1998]{koh98} Kohno, K.  1998, Ph.D Thesis, University of Tokyo

\bibitem[Kuno et al.\ 1997]{kun97} Kuno, N., Nakai, N., Nishiyama, K.,
    Sorai, K., Handa, T., \& Iga, T.  1997,
    in The Central Regions of Galaxy and Galaxies, ed. Y. Sofue
    (Dordrecht: Kluwer), in press

\bibitem[Lada 1992]{lad92} Lada, E.  1992, \apjl, 393, L25

\bibitem[Larson 1987]{lar87} Larson, R. B.  1987,
    in Starburst and Galaxy Evolution, ed. T. X. Thuan, T. Montmerle, \&
    J. Tran Thanh Van (Gif-sur-Yvette: Editions Frontieres), 467

\bibitem[Leep \& Dalgarno 1996]{lad96} Leep, S., \& Dalgarno, A.  1996, \aap, 306, L21

\bibitem[M\'arquez \& Moles 1993]{mam93} M\'arquez, I., \& Moles M. 1993,
    \aj, 105, 2090

\bibitem[Matsushita et al.\ 1998]{mat98} Matsushita, S., Kohno, K.,
   Vila-Vilar\'o, B., Tosaki, T., \& Kawabe, R.  1998, \apj, 495, 267 

\bibitem[Mauersberger \& Henkel 1989]{mau89} Mauersberger, R., \&
   Henkel, C. 1989, \aap, 223, 79

\bibitem[Mulchaey, Regan, \& Kundu 1997]{mrk97} Mulchaey, J. S.,
    Regan, M. W., \& Kundu, A.  1997, \apjs, 110, 229

\bibitem[Nguyen-Q-Rieu et al.\ 1992]{rie92} Nguyen-Q-Rieu,
    Jackson, J. M., Henkel, C., Truong-Bach, \& Mauersberger, R.
    1992, \apj, 399, 521


\bibitem[Oka et al.\ 1998]{oka98} Oka, T., Hasegawa, T., Hayashi, M., Handa, T.,
    Sakamoto, S., 1998, \apj, 493, 730 

\bibitem[Paglione et al.\ 1995]{pag95} Paglione, T. A. D.,
    Jackson, J. M., Ishizuki, S., \& Nguyen-Q-Rieu, 1995, \aj, 109, 1716

\bibitem[Paglione, Tosaki, \& Jackson 1995]{ptj95} Paglione, T. A. D.,
    Tosaki, T., \& Jackson, J. M.  1995, \apjl, 454, L117

\bibitem[Paglione, Jackson, \& Ishizuki 1997]{pag97} Paglione, T. A. D.,
    Jackson, J. M., \& Ishizuki, S.  1997, \apj, 484, 656

\bibitem[Paglione et al.\ 1998]{pag98} Paglione, T. A. D., Jackson, J. M.,
    Bolatto, A. D., \& Heyer, M. H.  1998, \apj, 493, 680

\bibitem[Peng et al.\ 1996]{pen96} Peng, R., Zhou, S., Whiteoak, J. B.,
    Lo, K. Y., \& Sutton, E. C.  1996, \apj, 470, 821

\bibitem[Pritchet 1977]{pri77} Pritchet, C.  1977, \apjs, 35, 397

\bibitem[Rieke, Lebofsky, \& Walker 1988]{rlw88} Rieke, G. H.,
    Lebofsky, M. J., \& Walker, C. E.  1988, \apj, 325, 679


\bibitem[Reynaud \& Downes 1997]{rad97} Reynaud, D., \& Downes, D.  1997,
    \aap, 319, 737 

\bibitem[Robert, Huntley, \& van Albada 1979]{rha79} Robert, W. W.,
    Huntley, J. M., \& van Albada, G. D.  1979, \apj, 233, 67

\bibitem[Rozas, Beckman, \& Knapen 1996]{roz96} Rozas, M., Beckman, J. E.,
    \& Knapen, J. H.  1996, \aap, 307, 735

\bibitem[Sage, Shore, \& Solomon 1990]{sss90} Sage, L J.,
    Shore, S. N., \& Solomon, P. M.  1990, \apj, 351, 422

\bibitem[Saikia et al.\ 1994]{sai94} Saikia, D. J., Pedlar, A.,
    Unger, S. W., \& Axon, D. J.  1994, \mnras, 270, 46

\bibitem[Sakamoto et al.\ 1995]{sak95} Sakamoto, K., Okumura, S. K.,
    Minezaki, T., Kobayashi, Y., \& Wada, K.  1995, \aj, 110, 2075

\bibitem[Sakamoto 1996]{sak96} Sakamoto, K. 1996, \apj, 471, 173

\bibitem[Sakamoto et al.\ 1998]{sak98} Sakamoto, K., Okumura, S. K.,
    Ishizuki, S., \& Scoville, N. Z.  1998, \apj, submitted.

\bibitem[Sakamoto 1997]{sak97} Sakamoto, S. 1997, \apjs, submitted.

\bibitem[RSA]{rsa} Sandage, A., \& Tammann, G.  1981, A Revised Sharpley-Ames
    Catalog of Bright Galaxies (Washington, DC: Carnegie Institution of 
    Washington)

\bibitem[Scoville et al.\ 1987]{sco87} Scoville, N. Z., Yun, M. S., Clemens, D. P.,
    Sanders, D. B., \& Waller, W. H.  1987, \apjs, 63, 821

\bibitem[Scoville et al.\ 1998]{sco98} Scoville, N. Z., Yun, 
    M. S., Armus, L, \& Ford, H.  1998, \apjl, 493, L63 

\bibitem[Shalabiea \& Greenberg 1996]{sag96} Shalabiea, O. M., 
    \& Greenberg, J. M.  1996, \aap, 307, 52

\bibitem[Shen \& Lo 1995]{sal95} Shen, J., \& Lo, K. Y.  1995,
    \apjl, 334, L99

\bibitem[Shioya \& Tosaki 1997]{sat97} Shioya, Y., \& Tosaki, T.
    1997, \apj, 484, 664

\bibitem[Shioya et al.\ 1998]{shi98} Shioya, Y., Tosaki, T., Ohyama, Y.,
    Murayama, T., Yamada, T., Ishizuki, S., \& Taniguchi, Y.  1998,
    \pasj, 50, 317

\bibitem[Solomon \& Scoville 1974]{sas74} Solomon, P. M., 
     \& Scoville, N. Z.  1974, \apjl, 187, L71

\bibitem[Solomon et al.\ 1987]{sol87} Solomon, P. M., Rivilo, A. R.,
     Barrett, J., and Yahil, A.  1987, \apj, 319, 730


\bibitem[Solomon et al.\ 1992]{sol92} Solomon, P. M.,
    Downes, D., \& Radford, S. J. E.  1992, \apj, 387, L55 



\bibitem[Tacconi et al.\ 1994]{tac94} Tacconi, L. J.,
    Genzel, R., Blietz, M., Cameron, M., Harris, A. I.,
    \& Madden, S.  1994, \apj, 426, L77 

\bibitem[Taniguchi et al.\ 1996]{tan96} Taniguchi, Y., Ohyama, Y.,
    Yamada, T., Mouri, H., \& Yoshida, M.  1996, \apj, 467, 215

\bibitem[Telesco, Dressel, \& Wolstencroft 1993]{tdw93} Telesco, C. M.,
    Dressel, L. L., \& Wolstencroft, R. D.  1993, \apj, 414, 120

\bibitem[Toomre 1964]{too64} Toomre, A. 1964, \apj, 139, 1217

\bibitem[Tully 1988]{tul88} Tully, R.  1988, Nearby Galaxies Catalog 
    (Cambridge: Cambridge Univ. Press)

\bibitem[Turner \& Ho 1983]{tah83} Turner, J. L., \& Ho, P. T. P. 1983, \apjl, 268, L79

\bibitem[Turner 1994]{tur94} Turner, J. L.  1994,
   in Mass-Transfer Induced Activity in Galaxies, ed. I. Shlosman
   (Cambridge: Cambridge Univ. Press), 91

\bibitem[Vila et al.\ 1990]{vil90} Vila M. B., Pedlar, A., Davies, R. D.,
    Hummel, E., \& Axon, D. J.  1990, \mnras, 242, 379

\bibitem[Walker, Lebofsky, Rieke 1988]{wlr88} Walker, C. E.,
    Lebofsky, M. J., \& Rieke, G. H.  1988, \apj, 325, 687


\bibitem[Wilson \& Ulvestad 1987]{wau87} Wilson, A. S., \& Ulvestad
    J. S.  1987, \apj, 319, 105

\bibitem[Wozniak et al.\ 1995]{woz95} Wozniak, H., Friedli, D.,
    Martinet, L., Martin, P., \& Bratschi, P.\ 1995, \aaps, 111, 115

\bibitem[Wright et al.\ 1993]{wri93} Wright, M. C. H., Ishizuki, S., Turner, J. L.,
    Ho, P. T. P., \& Lo, K. Y.  1993, \apj, 406, 470

\bibitem[Young et al.\ 1995]{you95} Young, J. S., Xie, S., Tacconi, L.,
    Knezek, P., Viscuso, P., Tacconi-Garman, L., Scoville, N., et al.,\ 
    1995, \apjs, 98, 219

\bibitem[Zhao et al.\ 1996]{zha96} Zhao, J. H., Anantharamaiah, K. R., Goss, W. M.,
    \& Viallefond, F.  1996, \apj, 472, 54

\end{thebibliography}
\end{document}